%% file: GPUTilings.tex
\documentclass[12pt,a4paper, twoside,reqno]{amsart}
\usepackage{amscd}
\usepackage{amsmath,amssymb,amsthm,mathrsfs}
\usepackage[margin=3cm]{geometry}
\usepackage{tikz}
\usepackage{enumitem}
\usetikzlibrary{calc}
\usetikzlibrary{matrix,arrows,decorations.pathmorphing,decorations.markings,calc}

\usepackage{subfig}
\usepackage{float}
\usepackage{parskip}
\usepackage{enumerate}
\usepackage{listings}
\usepackage{algorithm}
\usepackage[noend]{algpseudocode}

\makeatletter
\def\BState{\State\hskip-\ALG@thistlm}
\makeatother

\usepackage[font=small]{caption}

\theoremstyle{definition}

\newenvironment{SmallFigure}[1][]{\begin{figure}[#1]\vspace{-.2cm}}{\vspace{-.2cm}\end{figure}}

\newenvironment{PicFigure}[1][]{\begin{figure}[#1]\vspace{.4cm}}{\vspace{.4cm}\end{figure}}

\theoremstyle{remark}

\numberwithin{equation}{section}
\title{Random Tilings with the GPU}

\author{David Keating}
\address{D.K.: Department of Mathematics, University of California, Berkeley,
CA 94720, USA}
\email{dkeating@berkeley.edu}

\author{Ananth Sridhar}
\address{A.S.: Institut f{\"u}r Mathematik, MA 7-1, Technische Universit{\"a}t Berlin,
Stra{\ss}e des 17. Juni 136, 10623 Berlin, Germany}
\email{sridhar@math.tu-berlin.de}
\begin{document}

\maketitle

\begin{abstract}
We present GPU accelerated implementations of Markov chain algorithms to sample random tilings, dimers, and the six vertex model.
\end{abstract}

\section*{Introduction}

The study of tilings of planar regions by dominos, lozenges, and other shapes has a long history in mathematics, physics, and computer science. In the recent decades, the discovery of arctic circles, limit shapes, and other phenomena in random tilings of large regions has sparked a renewed excitement into their investigation. For a survey of these developments, see \cite{kenyondimers}.

A remarkable tool in the study of tilings has been the use of various computer programs to generate and visualize random tilings. A variety of powerful algorithms and techniques for sampling tilings have been developed and studied (see for example \cite{LRS,W,AR,SZ,PW}), and their standard implementations are widely available and well-utilized by researchers in the field.

On the other hand, the recent years have seen the advent of many new tools and techniques in high performance computing. In particular is the resurgence of parallel computing, driven by the fact that as speeds of single processors reach their physical limits, computers increasingly rely on multicore processors for performance and efficiency. So, while previously the sole purview of supercomputing, parallel computing is now-a-days essential to fully exploit modern computational power.

At the cutting-edge of multicore hardware is the graphics processing unit (GPU). Designed specifically for certain computations in 3D computer graphics, GPUs employ a massively-parallel architecture with up to thousands of limited-functionality processing cores working synchronously. Thanks to the popularity of video games, powerful GPUs are now common-place on nearly all personal computers,  as well as Playstations, XBoxes, and other devices. Utilizing these computational resources for tasks beyond computer graphics is a tantalizing prospect. Indeed, GPUs have proven to be well-suited for many other types of problems, and general purpose computing with the GPU (GPGPU) has become increasingly popular and successful in many fields. 

The main purpose of this note is to demonstrate the use of GPUs to generate random tilings. Our approach is based on the Glauber dynamics Markov chains where Markov moves are local ``flips" of the tiling. For parallelization, we consider non-local updates consisting of clusters of flips, which when chosen according to a domain decomposition, can be generated and executed independently and in parallel on GPU cores. We implemented the program with C++ and OpenCL, and have made our source code available \cite{github}.

The structure of the remainder of the paper is as follows: in the first section, we recall the basics of domino tilings and Markov chain algorithms for generating random tilings. We briefly discuss generalizations to some other models. In the second section, after reviewing some important aspects of graphics hardware architecture and programming, we explain our implementation of the Markov chain algorithm on the GPU. In the conclusion, we present the results of some experiments we conducted to test the program.

\noindent
\textbf{Acknowledgements:} We are very grateful to Nicolai Reshetikhin for many helpful discussions during the course of this work.  D.K. was supported by the NSF grant DMS-1601947. A.S. was supported by the NSF grant DMS-16011947 and by the DFG Collaborative Research Center TRR 109 ``Discretization in Geometry and Dynamics."

\section{Random Tilings}

\begin{figure}[b]
\subfloat[]{
\includegraphics[scale=.4]{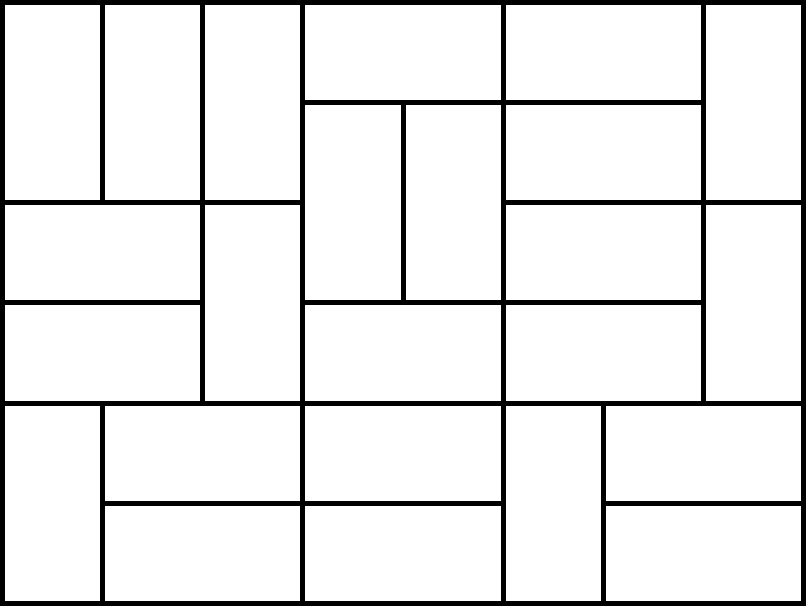}
} \; \; \;
\subfloat[]{
\includegraphics[scale=.4]{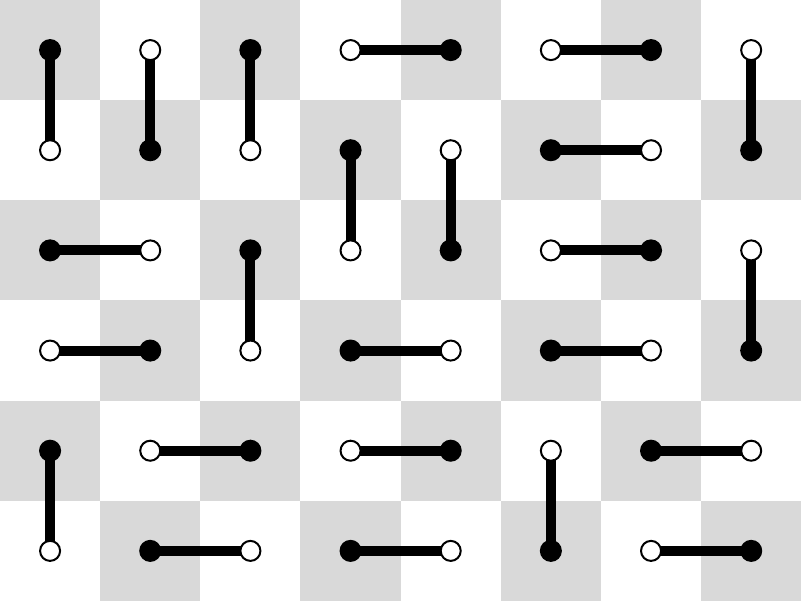}
} \; \; \;
\subfloat[]{
\includegraphics[scale=.4]{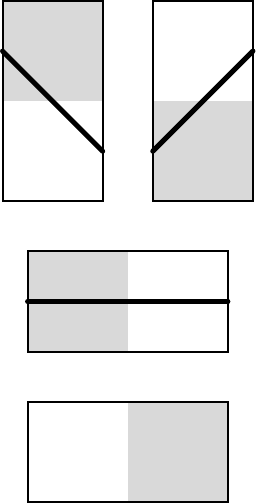}\; \;
\includegraphics[scale=.4]{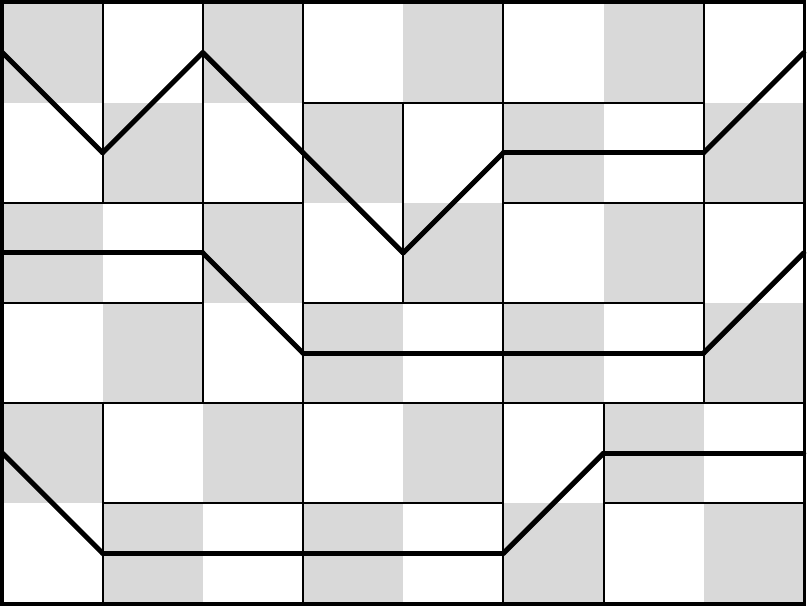} }
\caption{(A) A domino tiling of a rectangular domain. (B) The corresponding perfect matching of the dual graph. (C) The corresponding routing, obtained by drawing paths through dominos as shown.}
\label{fig:domtil}
\end{figure}

\subsection{Domino Tilings}\label{subsec:dom} By the $ \emph{square lattice} $ we mean the planar graph embedded in the Euclidean plane with vertices at coordinates $ (i,j) \in \mathbb{Z}^2 $ and edges joining all pairs of vertices unit distance apart. A \emph{domain} $ \mathcal{D} $ is the union of a finite set of faces of the square lattice. We assume that $ \mathcal{D} $ is simply-connected.

A \emph{domino} is the union of two square lattice faces that share an edge. A \emph{tiling} $\mathcal{T} $ of $ \mathcal{D} $ is a set of dominos whose interiors are disjoint and whose union is $ \mathcal{D} $. We denote by $ \Omega_\mathcal{D} $ the set of tilings of $ \mathcal{D}$. A domino tiling $ \mathcal{T} $ can equivalently be viewed as \emph{perfect matching} or \emph{dimer cover} $\mathcal{T}^*$ of the dual graph, or as a lattice routing as shown in Figure \ref{fig:domtil}.

\begin{figure}[t]
\subfloat[]{
\resizebox{.45\textwidth}{!} {%
\begin{tikzpicture} 
\node at (4,0){\includegraphics[scale=.75]{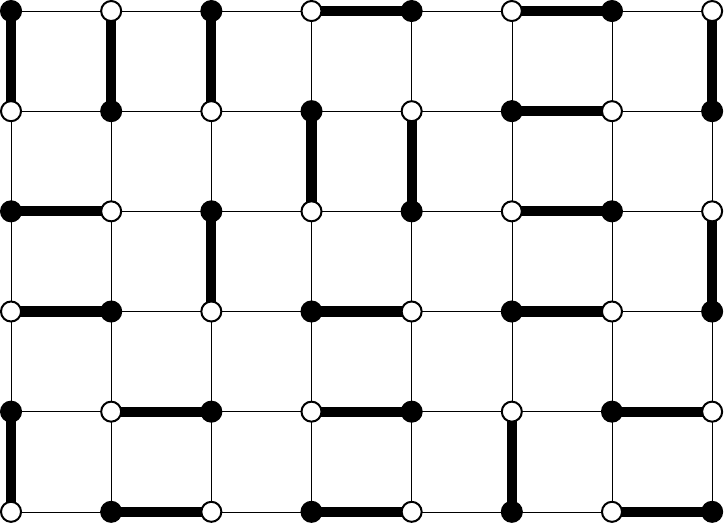}};
\node at (0,0){\includegraphics[scale=.75]{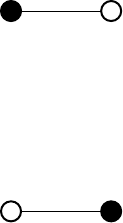}};

\node at  (1.75,-1.5) { \tiny $ 0 $ };
\node at  (2.5,-1.5) { \tiny $ -1 $ };
\node at  (3.25,-1.5) { \tiny $ 0 $ };
\node at  (4,-1.5) { \tiny $ -1 $ };
\node at  (4.75,-1.5) { \tiny $ 0 $ };
\node at  (5.5,-1.5) { \tiny $  3 $ };
\node at  (6.25,-1.5) { \tiny $ 4 $ };

\node at  (1.75,-.75) { \tiny $ 1 $ };
\node at  (2.5,-.75) { \tiny $ 2 $ };
\node at  (3.25,-.75) { \tiny $ 1 $ };
\node at  (4,-.75) { \tiny $ 2 $ };
\node at  (4.75,-.75) { \tiny $ 1 $ };
\node at  (5.5,-.75) { \tiny $ 2 $ };
\node at  (6.25,-.75) { \tiny $ 1 $ };

\node at  (1.75,0) { \tiny $ 4 $ };
\node at  (2.5,0) { \tiny $ 3 $ };
\node at  (3.25,0) { \tiny $ 0 $ };
\node at  (4,0) { \tiny $ -1 $ };
\node at  (4.75,0) { \tiny $ 0 $ };
\node at  (5.5,0) { \tiny $ -1 $ };
\node at  (6.25,0) { \tiny $ 0 $ };

\node at  (1.75,.75) { \tiny $ 1 $ };
\node at  (2.5,.75) { \tiny $ 2 $ };
\node at  (3.25,.75) { \tiny $ 1 $ };
\node at  (4,.75) { \tiny $ -2 $ };
\node at  (4.75,.75) { \tiny $ 1 $ };
\node at  (5.5,.75) { \tiny $ 2 $ };
\node at  (6.25,.75) { \tiny $ 1 $ };

\node at  (1.75,1.5) { \tiny $ 0 $ };
\node at  (2.5,1.5) { \tiny $ 3 $ };
\node at  (3.25,1.5) { \tiny $ 0 $ };
\node at  (4,1.5) { \tiny $ -1 $ };
\node at  (4.75,1.5) { \tiny $ 0 $ };
\node at  (5.5,1.5) { \tiny $ -1 $ };
\node at  (6.25,1.5) { \tiny $ 0 $ };

\draw[thin, decoration={markings,mark=at position 1 with {\arrow[scale=2]{>}}},
    postaction={decorate}] (-.75,.25) -- (-.75,1.25); \node at  (-1.05,.75) { \tiny $ +1 $ };
\draw[thin, decoration={markings,mark=at position 1 with {\arrow[scale=2]{>}}},
    postaction={decorate}] (-.75,-1.25) -- (-.75,-.25); \node at  (-1.05,-.75) { \tiny $ -1 $ };
\end{tikzpicture}
}} \; \; \; 
\subfloat[]{

\begin{tikzpicture}
\node at (-1.6,0){ \includegraphics[scale=.4]{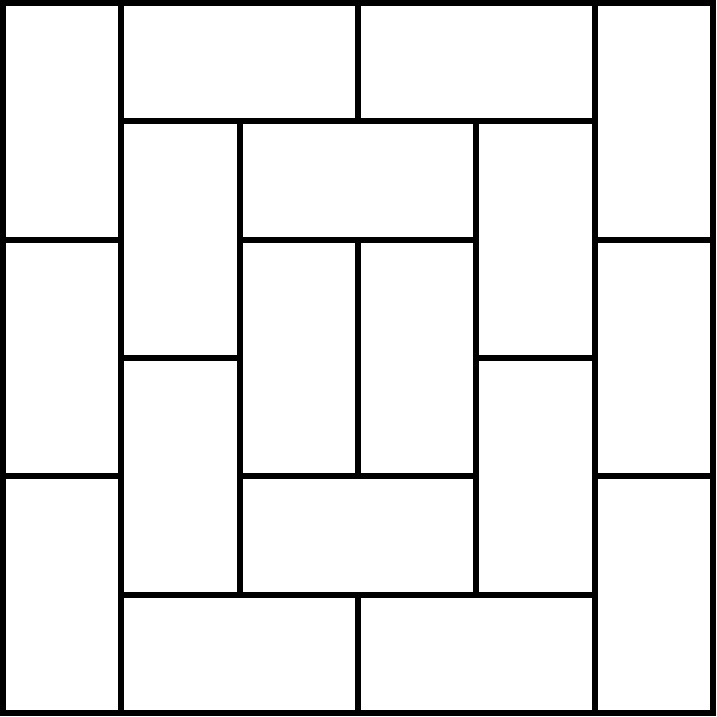}}; \;
\node at (1.6,0){ \includegraphics[scale=.4]{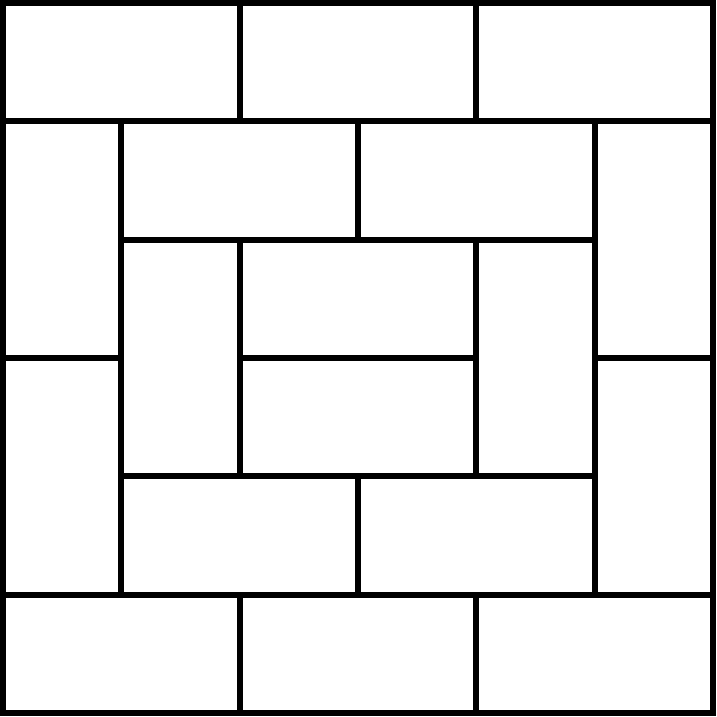}};
\end{tikzpicture}
}

\caption{(A) The height function is defined by first fixing the height at a reference face (say the bottom-most, left-most face) to zero and then propagating the height function across unmatched edges according to the rules at left. (B) The maximal and minimal tilings of a square domain.}
\label{fig:hf}
\end{figure}
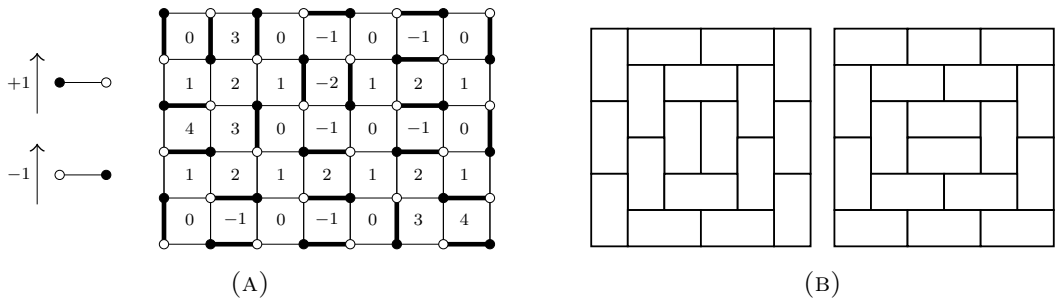

By identifying tilings with perfect matchings of a bipartite graph, each tiling $ \mathcal{T} $ can be associated with an integer valued \emph{height function} $ h_\mathcal{T} $ on vertices $ v \in \mathcal{D}$, according to the rules shown in Figure \ref{fig:hf}. Height functions induce a partial ordering on tilings, where $ \mathcal{T} < \mathcal{T}' $ if $ h_\mathcal{T}(v) < h_\mathcal{T}'(v) $ for all $ v \in \mathcal{D}$. Moreover, the point-wise maxima (or minima) of two height-functions also defines a height function and a corresponding tiling, which endows $ \Omega_\mathcal{D} $ with the structure of a distributive lattice. We denote by $ \mathcal{T}_{max} $ and $ \mathcal{T}_{min} $ the unique maximal and minimal elements of $ \Omega_\mathcal{D} $, see Figure \ref{fig:hf}.

We say a domain $ \mathcal{D} $ is \emph{tileable} if $ \Omega_\mathcal{D} $ is non-empty. An efficient algorithm due to \cite{Thurston} determines the tileability of a domain and returns a maximum (or minimum) tiling in time proportional to the size of the domain.

\subsubsection{Weights}\label{sec:weights}
It is often important to consider various probability measures on $ \Omega_\mathcal{D} $. Gibbs measures originate from models of statistical physics, in which dimer covers of the lattice correspond to bond configurations in crystals.

These measures are defined by a\emph{edge weights} that assign to each dual edge $ e $ a positive number $ w_ e $, and to each tiling $ \mathcal{T} $ a weight $ W(\mathcal{T}) = \prod_{e \in \mathcal{T^*}} w_e $. A tiling $ \mathcal{T} $ then has the probability
$$
P[\mathcal{T}] = \frac{1}{Z} \; W(\mathcal{T}), \hspace{30pt} Z = \sum_{\mathcal{T} \in \Omega } W(\mathcal{T} ).
$$ Equivalently, the Gibbs measure can be defined by \emph{volume weights} that assign to each vertex $ v $ a positive real number $ q_v $, and to each tiling $\mathcal{T} $ the weight $ W(\mathcal{T}) = \prod_{v \in \mathcal{D}} q_v^{h_\mathcal{T}(v)}. $ Choosing all weights to be 1 gives the uniform distribution on $ \Omega $.

\subsubsection{Moves on Tilings}
\begin{figure}[h]
\begin{tikzpicture}
\node at (-2.5,0){\includegraphics[scale=.4]{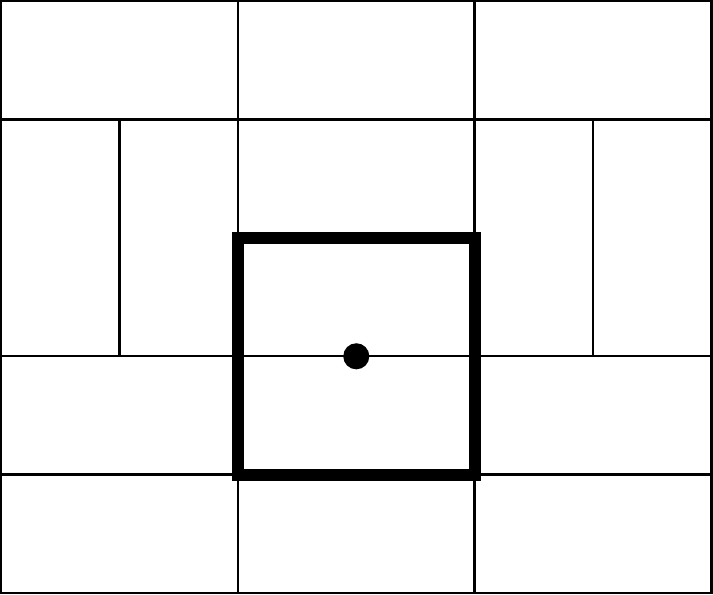}};
\node at (0,0) {$\longleftrightarrow$};
\node at (2.5,0){\includegraphics[scale=.4]{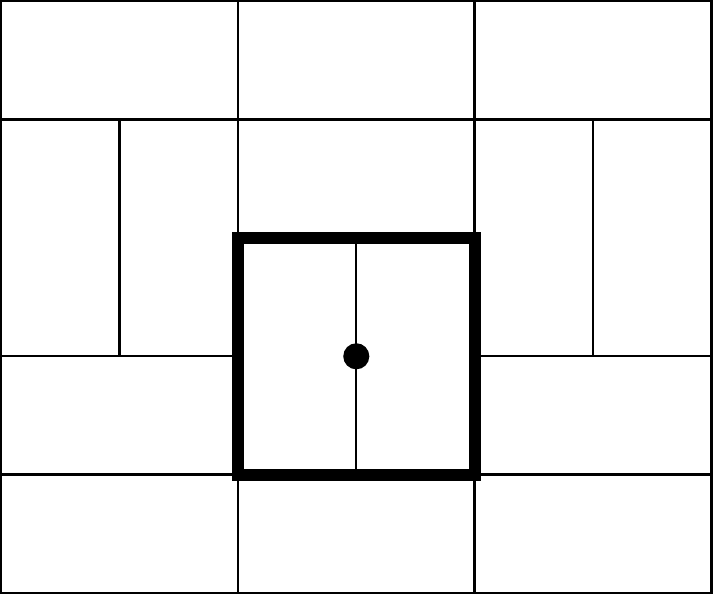}};
\end{tikzpicture}
\caption{An elementary rotation at a vertex.}
\label{fig:rot}
\end{figure}

We say a tiling $ \mathcal{T} $ is \emph{rotateable} at a vertex $ v $ if the faces adjacent to $ v $ are covered by two parallel dominos. An \emph{elementary rotation} at $ v $ rotates the dominos in place, as shown in Figure \ref{fig:rot}. More precisely, an up-rotation replaces two horizontal dominos by two vertical dominoes, and down rotation does the opposite. We denote by $ R_{\pm v}: \Omega_\mathcal{D} \rightarrow \Omega_\mathcal{D} $ the function that maps a tiling $ \mathcal{T} $ to the tiling obtained from $ \mathcal{T} $ by rotating up/down at $ v $ if possible. Here we adopt the convention of formally signing vertices to denote up and down rotations.

A theorem of \cite{Thurston} states that any two tilings $ \mathcal{T} $ and $ \mathcal{T}' $ of a domain $ \mathcal{D} $ are connected by a sequence of elementary rotations.

It is natural to consider simultaneous rotations on clusters of faces. Note that if two signed vertices $ v $ and $ v'$ are not adjacent, then $ R_{v} \circ R_{v'} = R_{v'} \circ R_v $. We call a subset of signed vertices $ S $  an \emph{admissible cluster} if no two of its vertices are adjacent. Then the cluster rotation $ R_S = \prod_{v\in S} R_v $ is well defined. In practice, a convenient class of admissible clusters is found by fixing a vertex coloring, and choosing subsets of vertices of the same color.

\subsubsection{Markov Chain on Tilings}
A random walk on $ \Omega_D $ is defined by an initial tiling $ \mathcal{T}^{(0)} $ and a sequence of random clusters $ \{ S_i \}_{i = 1 \cdots \infty} $, where the $n$th step is,
\begin{align} \label{mc}
\mathcal{T}^{(n)} = R^{(n)}(\mathcal{T} )\hspace{30pt} R^{(n)} = R_{S_n} \circ \cdots \circ R_{S_1}.
\end{align}
In other words, at each step, the Markov chain chooses a random cluster $ S $ and moves to $ R_S(\mathcal{T} )$. It is straightforward to check that this Markov chain is irreducible and aperiodic, from which it follows that in the limit $ n \rightarrow \infty $ the random tiling $ \mathcal{T}^{(n)} $ is uniformly distributed. 

In reality, the Markov chain is run for a finite but large time determined by the rate of convergence or \emph{mixing time}. Although mixing times of Markov chains on tilings have been studied by many, see for example \cite{LRS,W,LT}, and upper bounds rigorously established in many particular settings, very little is known in general. In practice, the mixing times can often be estimated empirically by heuristic techniques such as a self-consistent analysis of autocorrelation times. 

\subsubsection{Perfect Sampling with Coupling From the Past}
When statistical soundness is paramount, exact sampling can be accomplished using the \emph{coupling-from-the-past} algorithm \cite{PW}, which effectively simulates running the Markov chain for an infinite time. It works as follows: given a sequence of random clusters $ \{ S_i \}_{i = 1 \cdots \infty} $, define the backwards walk
\begin{align*}
\mathcal{T}^{(n)} = R^{(n)}(\mathcal{T} )\hspace{30pt} R^{(-n)} = R_{S_1} \circ R_{S_2} \circ \cdots \circ R_{S_n}
\end{align*}
Almost surely there exists an $ n $ for which $ | R^{(-n)}(\Omega_\mathcal{D}) | = 1 $ (and in fact for all earlier times $ m \geq n $). The Markov chain is then said to have \emph{collapsed} and the unique element in the range of $ R^{(-n)} $ is distributed according to the stationary distribution.

For Markov chains with large state spaces, checking for collapse can be impractical. However, in the case that the state space is partially ordered and the Markov moves monotone, as is the case for domino tilings, the state space collapses if and only if $ R^{(-n)}( \mathcal{T}_{max}) = R^{(-n)}(\mathcal{T}_{min}) $. Consequently, it is sufficient to check only the maximal and minimal states.

\subsection{Other Models}
The machinery described above for domino tilings can be generalized to a variety of other models. Let us briefly describe a few examples:

\subsubsection{Lozenge Tilings}
Lozenge tilings are the triangular-lattice analog of domino tilings, see Figure \ref{fig:lozenges}. Lozenge tilings correspond to dimers on the hexagonal lattice, whose bipartite structure allows the introduction of a height function at vertices (which is easy to visualize by imagining a tiling as a stack of cubes), which in turn induces a partial order and lattice structure on the set of tilings. The set is connected elementary rotations at vertices of the triangular lattice as shown below. Cluster rotations can be generated by three-coloring the triangular lattice.
\begin{SmallFigure}[H]
\subfloat[]{ \; \;
\begin{tikzpicture}
\clip(-.5,-.4) rectangle (.5,.4);
\begin{scope}[rotate=30.0,transform shape]
\node at (0,0){\includegraphics[scale=0.7]{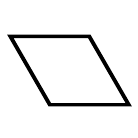}}; 
\end{scope}
\end{tikzpicture} \; \;
} \; \; \;
\subfloat[]{
\begin{tikzpicture}
\clip(-2.5,-1.25) rectangle (2.5,1.25);
\begin{scope}[shift={(-1.3,0)}, rotate=30.0,transform shape]
\node at (0,0){\includegraphics[scale=0.7]{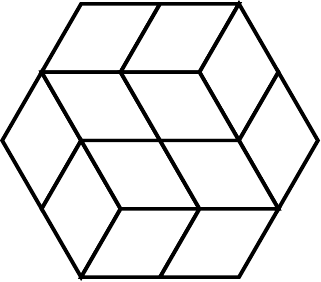}};
\end{scope}
\begin{scope}[shift={(1.3,0)}, rotate=30.0,transform shape]
\node at (0,0){\includegraphics[scale=0.7]{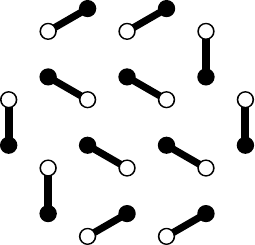}};
\end{scope}
\end{tikzpicture}
} \; \; \;
\subfloat[]{
\begin{tikzpicture}
\clip(-1.2,-.5) rectangle (1.2,.5);
\begin{scope}[shift={(-.8,0)}, rotate=30.0,transform shape]
\node at (0,0){\includegraphics[scale=.5]{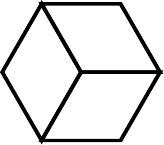}};
\end{scope}
\node at (0,0) {$\leftrightarrow$};
\begin{scope}[shift={(.8,0)}, rotate=30.0,transform shape]
\node at (0,0){\includegraphics[scale=.5]{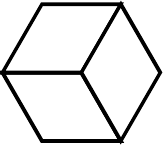}};
\end{scope}
;
\end{tikzpicture}
}
\caption{(A) A lozenge is a pair of equilateral triangles glued along a side. (B) A lozenge tiling and the corresponding matching on the hexagonal lattice. (C) An elementary rotation.} \label{fig:lozenges}
\end{SmallFigure}
\subsubsection{Bibone tilings}
Bibone tilings are the hexagonal-lattice analog of domino tilings, see Figure \ref{fig:bibones}. While they correspond to dimers on the triangular lattice, techniques for bipartite dimers do not apply; for example, bibone tilings do not admit height functions. Nonetheless, \cite{ClaireKenyon} showed the connectedness of the set of tilings under three types of elementary moves as shown in Figure \ref{fig:bibones}. For parallelization, we  consider clusters of the same type of move, with a different domain decomposition for each type.
\begin{SmallFigure}[H]
\subfloat[]{ 
\; \;
\includegraphics[scale=0.55]{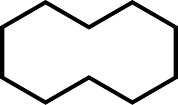}
\; \;
}
\subfloat[]{
\includegraphics[scale=0.55]{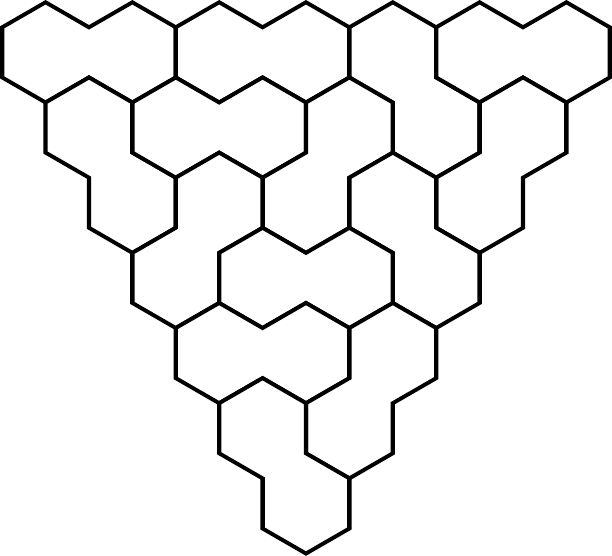}\; 
\begin{tikzpicture}[baseline]
\node at (0,1.5){\includegraphics[scale=0.55]{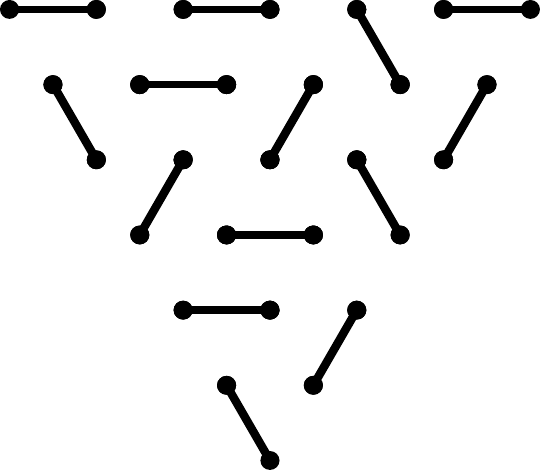}};
\end{tikzpicture}
} \; \;
\subfloat[]{
\shortstack{
\begin{tikzpicture}
\clip(-1.5,-.35) rectangle (1.5,.35);
\node at (-.8,0){\includegraphics[scale=.35]{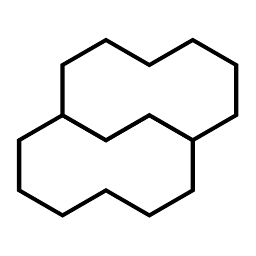}};
\node at (0,0) {$\leftrightarrow$};
\node at (.8,0){\includegraphics[scale=.35]{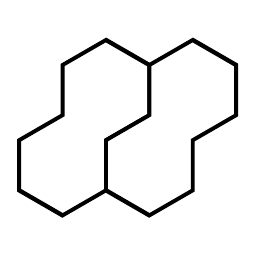}};
\end{tikzpicture} \\
\begin{tikzpicture}
\clip(-1.5,-.5) rectangle (1.5,.5);
\node at (-.8,0){\includegraphics[scale=.35]{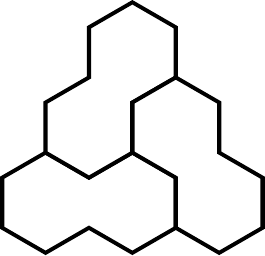}};
\node at (0,0) {$\leftrightarrow$};
\node at (.8,0){\includegraphics[scale=.35]{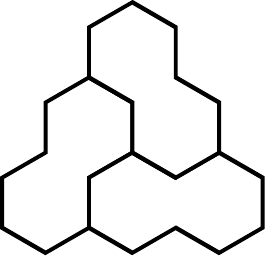}};
\end{tikzpicture} \\
\begin{tikzpicture}
\clip(-1.5,-.5) rectangle (1.5,.5);
\node at (-.8,0){\includegraphics[scale=.3]{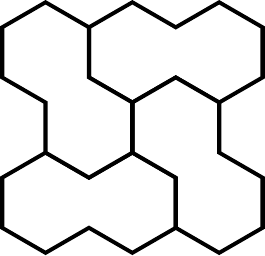}};
\node at (0,0) {$\leftrightarrow$};
\node at (.8,0){\includegraphics[scale=.3]{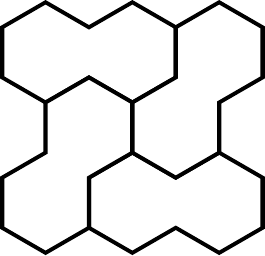}};
\end{tikzpicture}
}
}
\caption{(A) A bibone is a pair of hexagons glued along an edge. (B) A bibone tiling and the corresponding matching on the triangular lattice. (C) Up to orientation and reflection, there are three types of elementary moves.} \label{fig:bibones}
\end{SmallFigure}
\;

\subsubsection{Rectangle-triangle tilings} \label{sec:recttriangle}
Rectangle-triangle tilings were studied by B. Nienhuis, \cite{Nienhuis}. The tiles are isosceles triangles and rectangles with side lengths $ 1$ and $ \sqrt{3} $. We focus on tilings of domains of the triangular lattice. Like lozenge tilings, rectangle-triangle tilings can be visualized in 3D as stacks of half-cubes, which gives a partial ordering to the set of tilings. It is easy to check that the set of tilings is connected by an elementary move at vertices as shown in Figure \ref{fig:rtt}. In practice, we allow  many other local moves to improve the mixing rate.

\begin{SmallFigure}[h]
\subfloat[]{  \;
\begin{tikzpicture}
\clip(-1,-.6) rectangle (1,.6);
\begin{scope}[rotate=0.0,transform shape]
\node at (0,0){\includegraphics[scale=1.1]{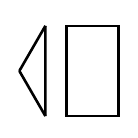}}; \; 
\end{scope}
\end{tikzpicture}
\;
} \; \;
\subfloat[]{
\begin{tikzpicture}
\clip(-1.5,-1.25) rectangle (1.5,1.25);
\begin{scope}[rotate=30.0,transform shape]
\node at (0,0){\includegraphics[scale=1.1]{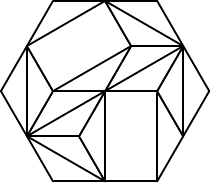}}; \; 
\end{scope}
\end{tikzpicture}
} \;  \; \;
\subfloat[]{ 
\shortstack{
\begin{tikzpicture} 
\clip(-1.2,-.5) rectangle (1.2,.5);
\draw[white] (-1.5,0)--(1.5,0);\begin{scope}[shift={(-.8,0)},rotate=30.0,transform shape]
\node at (0,0){\includegraphics[scale=.8]{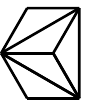}}; \end{scope}
\node at (0,0) {$\leftrightarrow$};\begin{scope}[shift={(.8,0)},rotate=30.0,transform shape]
\node at (0,0){\includegraphics[scale=.8]{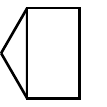}}; \end{scope}
\end{tikzpicture} 
}}
\caption{(A) Rectangle and triangle tiles. (B) A rectangle-triangle tiling. (C) Up to orientation and reflection, there is one elementary move.} \label{fig:rtt}
\end{SmallFigure}

Local weights can be introduced by assigning to each face of the triangular lattice a weight depending on the tiles that cover the face \cite{Nienhuis}. Up to orientation, there are four possibilities, as shown below: (A) covered by triangles, (B) covered by a rectangle and triangle, (C) covered by rectangles. These cases are assigned weights $ t$, $ c$, and $ r $ respectively; the weight of the tiling is given by the product of all weights of faces.
\begin{SmallFigure}[h]
\subfloat[][]{ \; \;
\begin{tikzpicture}[scale = .6]
\fill[lightgray] (0,0)--(.5,0.866025)--(1,0)--(0,0);
\draw (0,0)--(.5,0.866025)--(.5,-0.866025)--(0,0);
\draw (1,0)--(.5,0.866025)--(.5,-0.866025)--(1,0);
\end{tikzpicture} \; \; }
\subfloat[][]{ \; \;
\begin{tikzpicture}[scale = .6]
\fill[lightgray] (0,0)--(.5,0.866025)--(1,0)--(0,0);
\draw (0,0)--(.5,0.866025)--(.5,-0.866025)--(0,0);
\draw (1.5,0.866025)--(.5,0.866025)--(.5,-0.866025)--(1.5,-0.866025)--(1.5,0.866025);
\end{tikzpicture} \; \; }
\subfloat[][]{
\begin{tikzpicture}[scale = .6]
\fill[lightgray] (0,0)--(.5,0.866025)--(1,0)--(0,0);
\draw (.5,0.866025)--(-.5,0.866025)--(-.5,-0.866025)--(.5,-0.866025)--(.5,0.866025);
\draw (1.5,0.866025)--(.5,0.866025)--(.5,-0.866025)--(1.5,-0.866025)--(1.5,0.866025);
\end{tikzpicture}
\;
\begin{tikzpicture}[scale = .6]
\fill[lightgray] (-.5,-0.866025)--(0,0)--(.5,-0.866025)--(-.5,-0.866025);
\draw (.5,0.866025)--(-.5,0.866025)--(-.5,-0.866025)--(.5,-0.866025)--(.5,0.866025);
\end{tikzpicture} }
\end{SmallFigure}

\subsubsection{The six-vertex model}
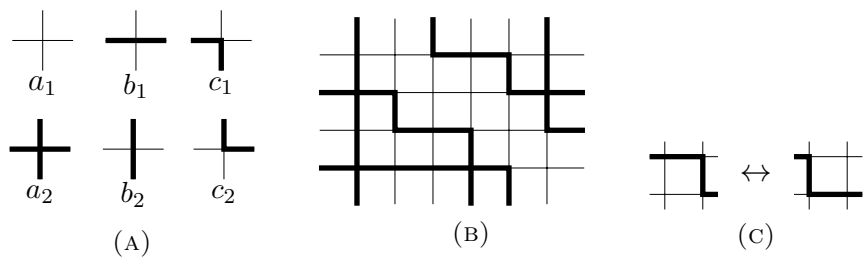
\begin{figure}[H]
\subfloat[]{
\shortstack{
\begin{tikzpicture}[scale=.5,baseline]
\draw[ultra thin] (-.8,0) -- (.8,0);
\draw[ultra thin] (0,-.8) -- (0,.8);
\node at (0,-1.2){$a_1$};
\end{tikzpicture} \; 
\begin{tikzpicture}[scale=.5,baseline]
\draw[ultra thin] (-.8,0) -- (.8,0);
\draw[ultra thin] (0,-.8) -- (0,.8);
\draw[line width=2pt](-.8,0)--(.8,0);
\node at (0,-1.2){\small{$b_1$}}; 
\end{tikzpicture}\;
\begin{tikzpicture}[scale=.5,baseline]
\draw[ultra thin] (-.8,0) -- (.8,0);
\draw[ultra thin] (0,-.8) -- (0,.8);
\draw[line width=2pt](0,-.8)--(0,0)--(-.8,0);
\node at (0,-1.2){\small{$c_1$}};
\end{tikzpicture} \\
\begin{tikzpicture}[scale=.5,baseline]
\draw[ultra thin] (-.8,0) -- (.8,0);
\draw[ultra thin] (0,-.8) -- (0,.8);
\draw[line width=2pt](-.8,0)--(.8,0);
\draw[line width=2pt](0,-.8)--(0,.8);
\node at (0,-1.2){\small{$a_2$}};
\end{tikzpicture} \; 
\begin{tikzpicture}[scale=.5,baseline]
\draw[ultra thin] (-.8,0) -- (.8,0);
\draw[ultra thin] (0,-.8) -- (0,.8);
\draw[line width=2pt](0,-.8)--(0,.8);
\node at (0,-1.2){\small{$b_2$}};
\end{tikzpicture} \; 
\begin{tikzpicture}[scale=.5,baseline]
\draw[ultra thin] (-.8,0) -- (.8,0);
\draw[ultra thin] (0,-.8) -- (0,.8);
\draw[line width=2pt](.8,0)--(0,0)--(0,.8);
\node at (0,-1.2){\small{$c_2$}};
\end{tikzpicture}}
}  \; \;
\subfloat[]{
\begin{tikzpicture}[scale = 1,baseline]\begin{scope}[shift={(0,.75)}]
\draw[step=.5 cm, ultra thin] (-1.99,-1.48) grid (1.49,.98);
\draw[line width=2pt] (-1.5,-1.5)--(-1.5,0)--(-2,0); \draw[line width=2pt] (1.5,0)--(.5,0)--(.5,.5)--(-.5,.5)--(-.5,1);
\draw[line width=2pt] (.5,-1.5)--(.5,-1)--(-2,-1); \draw[line width=2pt] (1.5,-.5)--(1,-.5)--(1,1);
\draw[line width=2pt] (0,-1.5)--(0,-.5)--(-1,-.5)--(-1,0)--(-1.5,0)--(-1.5,1);
\end{scope}
\end{tikzpicture} \; \;
} \subfloat[]{
\begin{tikzpicture}[baseline]
\begin{scope}[shift={(0,-.6)}]
\begin{scope}[shift={(-1.2,0)}]
\draw[step=.5 cm, very thin] (-0.2,-0.2) grid (.7,.7);
\draw[line width=2pt] (.7,0)--(.5,0)--(.5,.5)--(-.2,.5);
\end{scope}
\node at (0,.25) {$\leftrightarrow$};
\begin{scope}[shift={(.7,0)}]
\draw[step=.5 cm, very thin] (-0.2,-0.2) grid (.7,.7);
\draw[line width=2pt] (.7,0)--(.5,0)--(0,0)--(0,.5)--(-.2,.5);
\end{scope}
\end{scope}
\end{tikzpicture}
}
\caption{(A) The six vertex types with six weights. (B) A six-vertex configuration. (C) A local move.} \label{fig:sixvertex}
\end{figure}

The six-vertex model is defined on a domain $\mathcal{D} $  of the square lattice. A configuration $ S$ of the model is an assignment of ``occupied" or ``unoccupied" to each edge in $\mathcal{D}$ that satisfies the condition that at every vertex $ v $ in the interior of $\mathcal{D} $, the edges adjacent to $ v $ must be one of six local configurations shown in Figure \ref{fig:sixvertex}. A boundary condition for the six-vertex model fixes the state of edges intersecting the boundary of $ \mathcal{D}$.

\begin{figure}[h]
\begin{tikzpicture}[scale=.5]
\draw[very thin] (-.5,0)--(-.5,2)--(.5,2)--(.5,0)--(-.5,0);
\draw[ultra thick] (-.5,1)--(.5,1);
\draw[decoration={markings,mark=at position 1 with {\arrow[scale=2]{>}}},
    postaction={decorate}] (-.8,.4) -- (-.8,1.6);
\node at (-1.3,1){\tiny{$+1$}};
\draw[very thin] (-1,-1.5)--(-1,-.5)--(1,-.5)--(1,-1.5)--(-1,-1.5);
\draw[ultra thick] (0,-1.5)--(0,-.5);
\draw[decoration={markings,mark=at position 1 with {\arrow[scale=2]{>}}},
    postaction={decorate}] (-.6,-1.8) -- (.6,-1.8);
\node at (0,-2.2){\tiny{$+1$}};
\end{tikzpicture}
\begin{tikzpicture}[scale=.5]
\draw[very thin] (-.5,0)--(-.5,2)--(.5,2)--(.5,0)--(-.5,0);
\draw[very thin] (-.5,1)--(.5,1);
\draw[decoration={markings,mark=at position 1 with {\arrow[scale=2]{>}}},
    postaction={decorate}] (-.8,.4) -- (-.8,1.6);
\node at (-1.3,1){\tiny{$-1$}};
\draw[very thin] (-1,-1.5)--(-1,-.5)--(1,-.5)--(1,-1.5)--(-1,-1.5);
\draw[very thin] (0,-1.5)--(0,-.5);
\draw[decoration={markings,mark=at position 1 with {\arrow[scale=2]{>}}},
    postaction={decorate}] (-.6,-1.8) -- (.6,-1.8);
\node at (0,-2.2){\tiny{$-1$}};
\end{tikzpicture} \hspace{30pt}
\begin{tikzpicture}[scale=1]
\node at (-1.75,-1.25){\tiny{0}};
\node at (-1.25,-1.25){\tiny{1}};
\node at (-.75,  -1.25){\tiny{0}};
\node at (-.25,  -1.25){\tiny{-1}};
\node at (.25,   -1.25){\tiny{0}};
\node at (.75,   -1.25){\tiny{1}};
\node at (1.25, -1.25){\tiny{0}};

\node at (-1.75,-.75){\tiny{1}};
\node at (-1.25,-.75){\tiny{2}};
\node at (-.75,-.75){\tiny{1}};
\node at (-.25,-.75){\tiny{0}};
\node at (.25,-.75){\tiny{1}};
\node at (.75, -.75){\tiny{0}};
\node at (1.25, -.75){\tiny{-1}};

\node at (-1.75,-.25){\tiny{0}};
\node at (-1.25,-.25){\tiny{1}};
\node at (-.75,-.25){\tiny{2}};
\node at (-.25,-.25){\tiny{1}};
\node at (.25,-.25){\tiny{0}};
\node at (.75, -.25){\tiny{-1}};
\node at (1.25, -.25){\tiny{0}};

\node at (-1.75,.25){\tiny{1}};
\node at (-1.25,.25){\tiny{2}};
\node at (-.75,.25){\tiny{1}};
\node at (-.25,.25){\tiny{0}};
\node at (.25,.25){\tiny{-1}};
\node at (.75, .25){\tiny{0}};
\node at (1.25, .25){\tiny{1}};

\node at (-1.75,.75){\tiny{0}};
\node at (-1.25,.75){\tiny{1}};
\node at (-.75,.75){\tiny{0}};
\node at (-.25,.75){\tiny{1}};
\node at (.25,.75){\tiny{0}};
\node at (.75, .75){\tiny{-1}};
\node at (1.25, .75){\tiny{0}};
\draw[step=.5 cm, ultra thin] (-1.99,-1.48) grid (1.49,.98);
\draw[line width=2pt] (-1.5,-1.5)--(-1.5,0)--(-2,0); \draw[line width=2pt] (1.5,0)--(.5,0)--(.5,.5)--(-.5,.5)--(-.5,1);
\draw[line width=2pt] (.5,-1.5)--(.5,-1)--(-2,-1); \draw[line width=2pt] (1.5,-.5)--(1,-.5)--(1,1);
\draw[line width=2pt] (0,-1.5)--(0,-.5)--(-1,-.5)--(-1,0)--(-1.5,0)--(-1.5,1);
\end{tikzpicture}
\caption{The height function for the six-vertex model.} \label{fig:sixvertexh}
\end{figure}
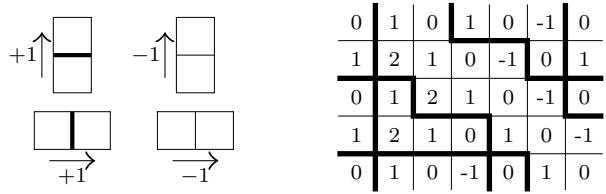

Six-vertex configurations correspond bijectively to height functions, as shown in Figure \ref{fig:sixvertexh}. This endows the set of configurations  with a partial ordering and the structure of a distributive lattice.

Gibbs measures are defined by assigning positive real weights to each of the six vertex types. The weight of a configuration $ S $ is $ W(S) = \prod_{v \in \mathcal{D}} w(v,S) $ where $ w(v,S) $ is the weight of the vertex $ v $ in the configuration $ S $ depending on the vertex type. The probability of $ S$ is then
\begin{align*}
P[S]= \frac{1}{Z} W(S) \hspace{20pt} Z = \sum_{S \in \Omega} W(S)
\end{align*}
The weights $ a = 1, b = 1, c= 1 $ give the uniform distribution on configurations.

With fixed boundary conditions, the set of configurations is connected by elementary $c$-flips at faces, as shown in Figure \ref{fig:sixvertex}. In general, variations of $ c$-flips for boundary faces are also required. For parallelization, cluster moves can be generated by coloring faces according to both the parity of the $x$- and $y$- coordinate of the face.

It can be checked that the local moves preserve the partial order only if the weights satisfy $ a \leq c $ and $ b \leq c $; this means that coupling-from-the-past can only be used for these weights, although the standard Markov chain sampling can be used for all weights. For the weights $ (a = 1, b = 1, c = 1) $, a much faster program is possible due to the combinatorial nature of the model, although we do not pursue it here.

\section{Implementation}

\subsection{Graphics Processing Units}
Certain properties of graphics hardware have great significance in the design and implementation of efficient algorithms for the GPU. Let us briefly mention a few important aspects, and refer the reader to \cite{HenPat} for details.

\subsubsection{GPU Design and Architecture}
Many tasks in 3D computer graphics involve carrying out an identical sequence of operations independently on a large set of input primitives. Rendering a surface for instance could require: for each pixel drawn to the screen, computing the local surface normal, computing directions to the viewer and light sources, and finally computing the color according to some lighting and shading model. These types of tasks are often called \emph{data parallel}, and the routine executed on each input called a shader or \emph{kernel}.

GPUs are designed specifically for data parallel tasks. At the hardware level, a typical GPU consists of a number of \emph{compute units}, each containing an instruction block, shared memory caches, and a number of \emph{processing elements} that carry out arithmetic computations. A parallel computation is organized by issuing a \emph{thread} for each kernel execution instance; the threads are divided into blocks and distributed to compute units for execution. Each compute unit executes the kernel in \emph{single instruction multiple data} (SIMD) fashion, by issuing at each time-step the same kernel instruction to every processing element to carry out in a different thread.

For tasks well-suited to this architecture, GPUs can perform up to hundreds of times faster than conventional CPUs.

Performance can drastically suffer, however, when programs are not tuned for the graphics hardware. Due to the SIMD execution model, GPU performance can be severely affected by \emph{branch divergences}, which occur when conditional expressions in the kernel cause different threads to follow different execution paths. Furthermore, compared to conventional processors, GPUs have little hardware dedicated to speed-up the execution of sequential code. In particular, GPU cores have relatively smaller memory caches and limited bandwidth for memory access. While GPUs hide memory-access latency by multithreading, their performance is nonetheless sensitive to the pattern of memory-access and organization of data in memory.

\subsubsection{The OpenCL Framework}
In early years, general purpose computing with GPUs required reformulating a task in terms of graphics primitives and operations. Since then, GPUs have evolved into highly flexible and easily programmable processors. Frameworks such as NVidia's Compute Unified Device Architecture (CUDA) and the Open Computing Language (OpenCL), define C-type kernel programming languages and provide C/C++ libraries to access GPU devices.

For this work, we chose to use OpenCL. 

In the OpenCL framework, a computing platform consists of a \emph{host} (typically the CPU) and any number of \emph{devices} (typically GPUs). Kernel source code is loaded and compiled for devices by OpenCL at run-time. The host issues commands to the device, such as kernel executions and data transfers, via a command queue. Before launch, the set of kernel execution threads is arranged by the user into a grid of \emph{work-items}. The grid coordinates of a work-item can be accessed from within the kernel program and used to identify the thread.  The grid is further divided into \emph{work-groups}, each of which is executed on one compute unit of the GPU, and whose work-items can communicate via shared local memory and can synchronize with barriers.

A typical OpenCL host program first initializes the platform, then loads and compiles the kernel source, transfers relevant data to the device, sets up and launches the kernel on the device, and finally reads back the output data from the device. 

\subsection{Random Domino Tilings with the GPU}
Having laid the necessary foundations, let us now explain our implementation.

To each vertex $v$ we associate an integer \emph{state} representing the tiling of the faces adjacent to $ v $ as follows: enumerating the edges adjacent to $v$ in the order North, South, East, West, the tilestate $s_v$ is defined as
\begin{align*}
s_v = e_0+2 \;e_1+4\; e_2+8\;e_3
\end{align*}
where $e_i$ is 1 if  a domino of $ \mathcal{T} $  crosses edge $i$ and 0 otherwise. A vertex $ v $ is rotateable if its state $ s_v = 12 $ or $ s_v = 3$. 

\begin{figure}[b]
\begin{tikzpicture}
\node at (-2,0){\includegraphics[scale=.2]{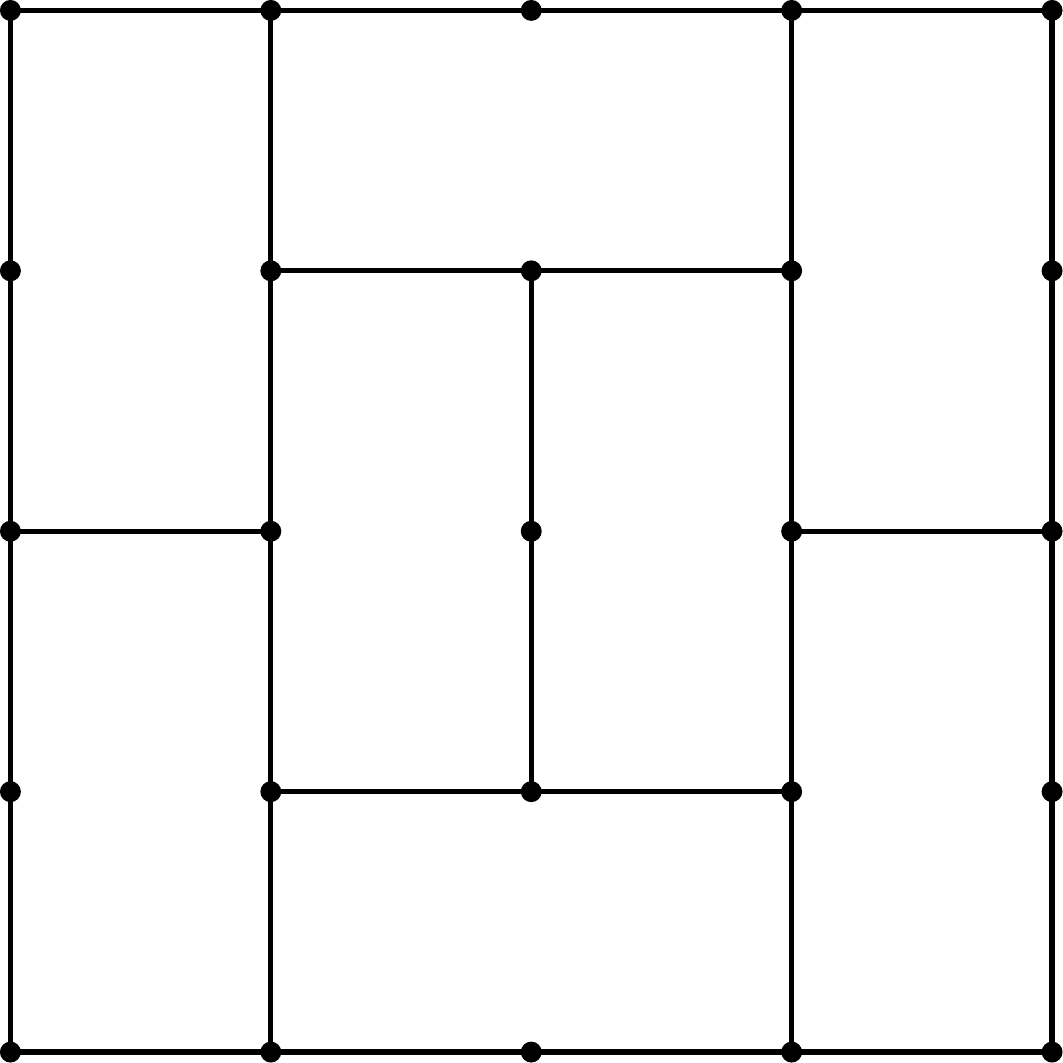}};

\node at (2,0) {\tiny
$
\left(
\begin{array}{ccccc}
 0 & 0 & 2 & 0 & 0 \\
 8 & 4 & 1 & 8 & 4 \\
 0 & 8 & 12 & 4 & 0 \\
 8 & 4 & 2 & 8 & 4 \\
 0 & 0 & 1 & 0 & 0 \\
\end{array}
\right)$};
\end{tikzpicture}
\caption{A tiling and its tilestate (without zero padding).}
\label{fig:tilestate}
\end{figure}

A tiling is represented by the state of every vertex of the domain $ \mathcal{D} $, as in Figure \ref{fig:tilestate}. Assuming the domain $ \mathcal{D} $ is contained in   an $ N \times N $ square domain in the first quadrant of the Euclidean plane, we store a tiling $ \mathcal{T} $ as an array $ N \times N $ array $ T $, where the $ (i,j) $th entry of $ T $ is the state of the vertex with coordinate $ (i,j) $. 

To optimize kernel occupancy and memory transfer speed, we divide the tiling into the sub-arrays $ T_b $, $ T_w $  of black and white vertices, and store each contiguously in the GPU device memory. For simplicity, we assume $ N $ is even, so that for example the $ (i,j)$th component of the array $ T_b $ corresponds to the state of the vertex $ (i,2 j+ i \text{ mod } 2) $ in $ \mathcal{D} $. Moreover, to avoid special cases at the boundary we assume all arrays are sufficiently zero-padded. 

In our implementation, the primitives of GPU operations are the vertices of the domain $ \mathcal{D} $.  We define a kernel function that rotates a vertex $ v $ with some fixed probability, and kernel functions that check the neighbors of $ v $ and recomputes its state. More precisely, 
\begin{itemize}[leftmargin=15pt,itemsep=-2pt]
\item The kernel \emph{Rotate} takes as a parameter a tiling sub-array $ T $. The work-item $ (i,j) $ first generates a pseudo-random number, and then with fixed probability attempts to either rotate up or down the state $ T[i,j] $. The state can be rotated up if $ T[i,j] = 3 $ with rotating accomplished by setting $ T[i,j] =  12$, and similarly for down-rotation. 
\item The kernels \emph{UpdateBlack} and \emph{UpdateWhite} take as parameters both sub-arrays $ T_b, T_w $. The work-item $ (i,j) $ of \emph{UpdateBlack} recomputes the state of the vertex with index $ (i,j)$ in $ T_b $ in terms of the neighboring vertices in $ T_w $. This can be done efficiently with bitwise operations, as follows
\begin{align*}
T_b[i,j] =& \frac{1}{2} (T_w[i-1,j]\; \& \;2) \; +  \; 2 (T_w[i+1,j] \; \& \;1) + \\ &  \frac{1}{2} (T_w[i, j + i \text{ mod } 2 - 1] \; \& \; 8) \; + 2(T_w[i,j + i \text{ mod } 2]\; \& \;4)
\end{align*}
\end{itemize}
The \emph{UpdateWhite} kernel is defined similarly.

Other kernels used to generate pseudo-random numbers are discussed in Section \ref{subsec:GPUPRNG}.

A random cluster rotation is accomplished by launching \emph{Rotate} on all black vertices and \emph{Update} on all white vertices, or \emph{Rotate} on all white vertices and \emph{Update} on all black vertices. We define the following host functions:
\begin{itemize}[leftmargin=15pt,itemsep=-2pt]
\item The function \emph{RandomWalk} takes as parameters a tiling $ \mathcal{T} $, a random number seed $ s $, and a natural number $ nSteps $. The function first divides $ \mathcal{T} $ into the two sub-arrays $ T_b $ and $ T_w $ and loads them to the GPU memory. It then seeds all pseudo-random number generators with $ s $. Then, looping $ nSteps $ times, with equal probability it either runs \emph{Rotate} with all black vertices and \emph{Update} on all white vertices, or vice versa. Finally, after the loop is complete, it reads back $ T_b $ and $ T_w $, and recombines the two sub-arrays into the tiling $ T$.

\item The function \emph{DominoTilerCFTP} generates a random tiling using the coupling-from-the-past algorithm. The pseudo-code is:
\begin{itemize}[leftmargin=20pt, itemsep=-6 pt]
\item[] {\bf DominoTilerCFTP:}
\item[] Compute the extremal tilings $ \mathcal{T}_{max} $ and $ \mathcal{T}_{min} $.
\item[] Initialize a list $ seeds $ with a random real number.
\item[] Repeat:
\item[] \begin{itemize}[ itemsep=-6 pt]
\item[] Initialize $ \mathcal{T}_{top} = \mathcal{T}_{max} $
\item[] Initialize $ \mathcal{T}_{bottom} = \mathcal{T}_{min} $
\item[] For \emph{i} = 1 to length of \emph{seeds}:  
\item[] \hspace{10pt} Set $\mathcal{T}_{top} =  RandomWalk(\mathcal{T}_{top}, seeds_i,  2^i  ) $
\item[] \hspace{10pt} Set $\mathcal{T}_{bottom} =  RandomWalk(\mathcal{T}_{bottom}, seeds_i,  2^i ) $
\item[] If $ \mathcal{T}_{top} = \mathcal{T}_{bottom} $, return $ \mathcal{T}_{bottom} $.
\item[] Else, push a new random number to the beginning of \emph{seeds}.
\end{itemize}
\end{itemize}
\end{itemize}

\subsubsection{Pseudo-random Number Generators}\label{subsec:GPUPRNG}
Random number generators for massively parallel computing is an active area of research. Most applications require a stream of pseudo-random numbers for each thread, with good statistical properties both within each stream and across different streams.  The statistical independence is crucial in our application to ensure that the set of tilings is connected by the cluster moves generated by the \emph{Rotate} kernel.
 
For our implementation, we adopted a variant of the well-known \emph{Mersenne Twister} pseudo-random number generator known as TinyMT \cite{TinyMT}. Although compared to the Mersenne Twister, the period of TinyMT is relatively small ($ 2^{127} $), the TinyMT admits a large number (up to $ 2^{48} $) of internal parameter values that ensure statistical independence (see \cite{TinyMT}) of generators with different parameter values. For tilings with fewer than $ 2^{48} $ vertices, it is convenient to simply instantiate a TinyMT generator with a unique parameter value for each vertex of the domain.

\section{Conclusion}
We tested our implementation using a few different graphics cards: an Intel Integrated HD 510, an NVidia Tesla P100, and an AMD Radeon Pro 555. Although a rigorous performance comparison of GPU algorithms and CPU algorithms can be a delicate matter (see \cite{D10}), we nevertheless compared our implementation with standard CPU algorithms, see Figure \ref{fig:cpugpu}. We found that for smaller domains, the processing time is dominated by memory transfer time and other overheads, and the CPU is often faster. For larger domains, the parallelism pays off and we found significant speed-ups, of at least an order of magnitude depending on the quality of the graphics hardware.

\begin{PicFigure}[h]
\subfloat[Subfigure 1][]{
\includegraphics[width=0.45\textwidth]{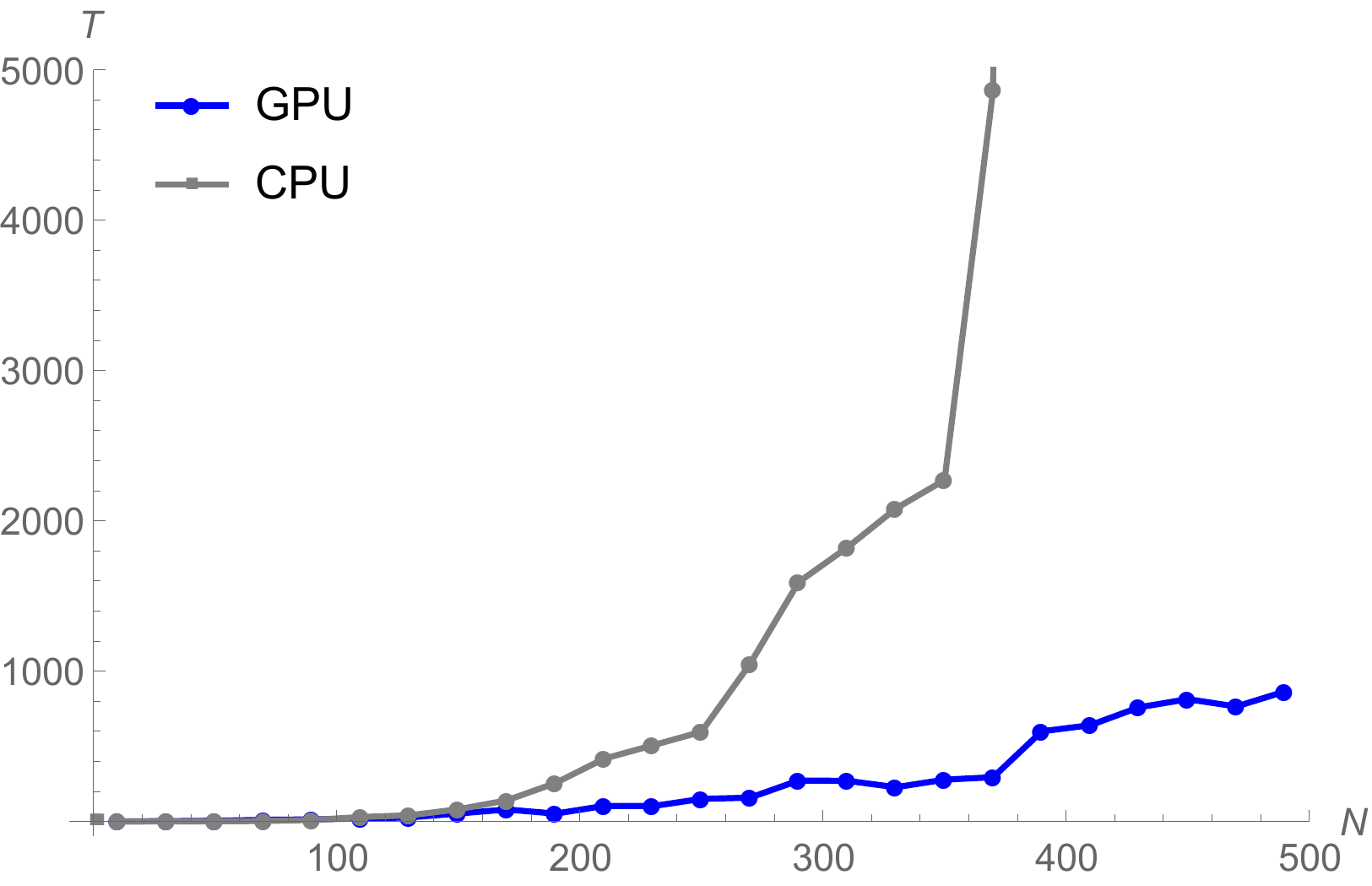}
\label{fig:pa}} \; \; \; 
\subfloat[Subfigure 2][]{
\includegraphics[width=0.45\textwidth]{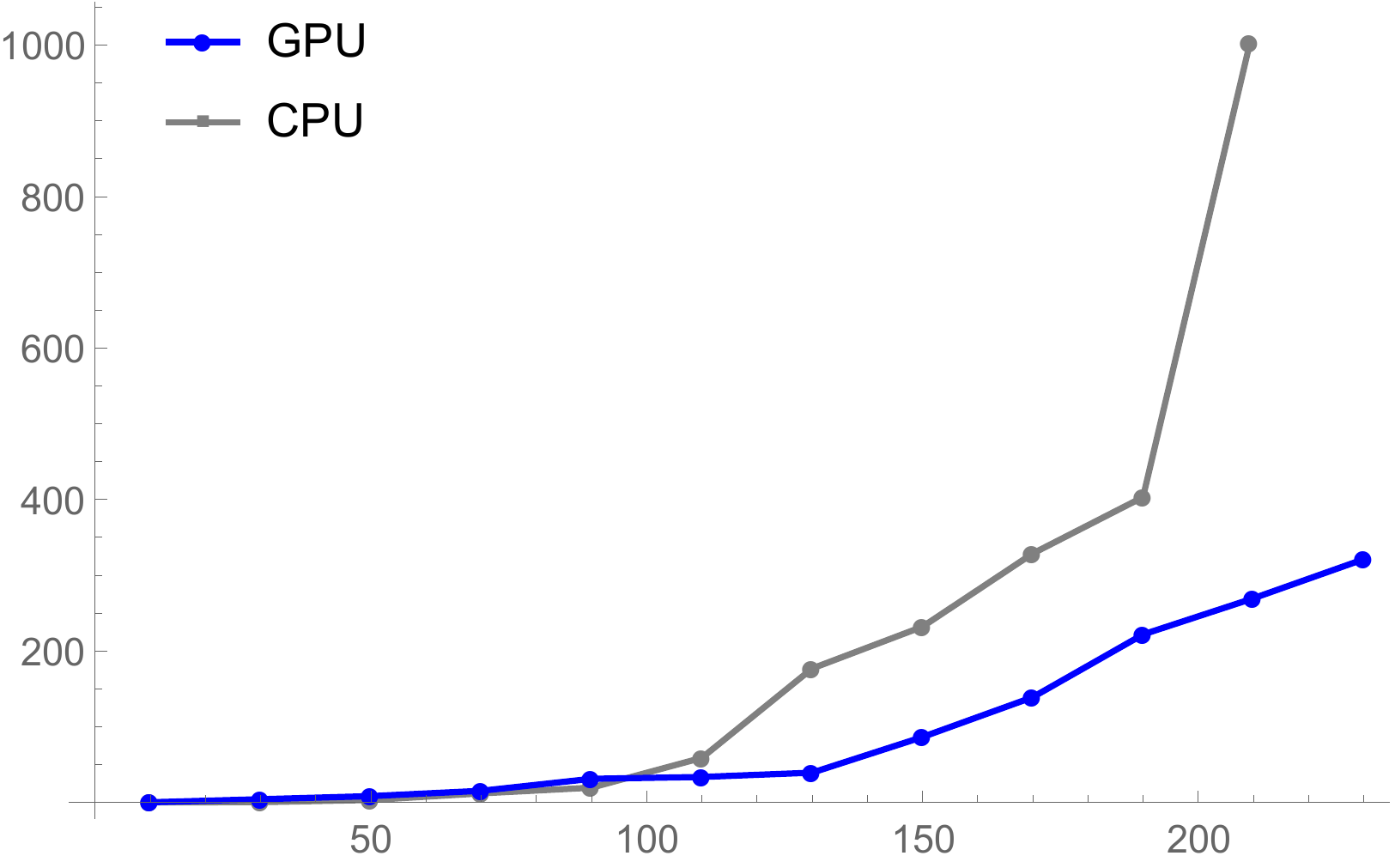}
\label{fig:pb}}
\caption{(A) The time $ T $ in seconds to generate, with coupling-from-the-past, a random configuration of the six-vertex model on an $N\times N$ sized domain with domain wall boundary conditions and weights $ (a,b,c) = (1,1,1) $. We used an Nvidia Tesla P100 GPU and a 2.2 GHz Intel Xeon E5 CPU. (B) The time to generate a random domino tiling of an $ N \times N $ square. Here we used a laptop with an Intel 510 Integrated GPU and 2.10 GHz Pentium CPU.}\label{fig:cpugpu}
\end{PicFigure}

There are several avenues for improvement and generalization that we leave for future work.

Firstly, our implementation leaves some opportunities for optimization, which we have forgone in favor of simplicity and flexibility. On the other hand, the fundamental limiting factor of GPU performance, particularly for large domains, is the number of available processing cores. Therefore, a natural next step is the use of multiple GPUs in parallel with message passing, which has proven successful in other applications.

More broadly, various other algorithms have been developed for generating tilings and could be adapted effectively for the GPU. Among the earliest were shuffling algorithms for exact sampling tilings of the Aztec diamond. Shuffling-type algorithms have since been generalized to many other settings \cite{Cedric}.

Finally, there are many models closely related to tilings whose simulations are of great interest. In particular, considerable recent work has focused on certain stochastic processes, including the stochastic six-vertex model, exclusion processes, Schur processes, and others. While these models in a sense arise as special limits of dimer and vertex models, it is clear that their simulation could benefit from a different approach.
\vfill
\pagebreak
\section{Examples}
\begin{figure}[h]
\begin{tikzpicture}
\node at (-2,4){ \includegraphics[scale = .8]{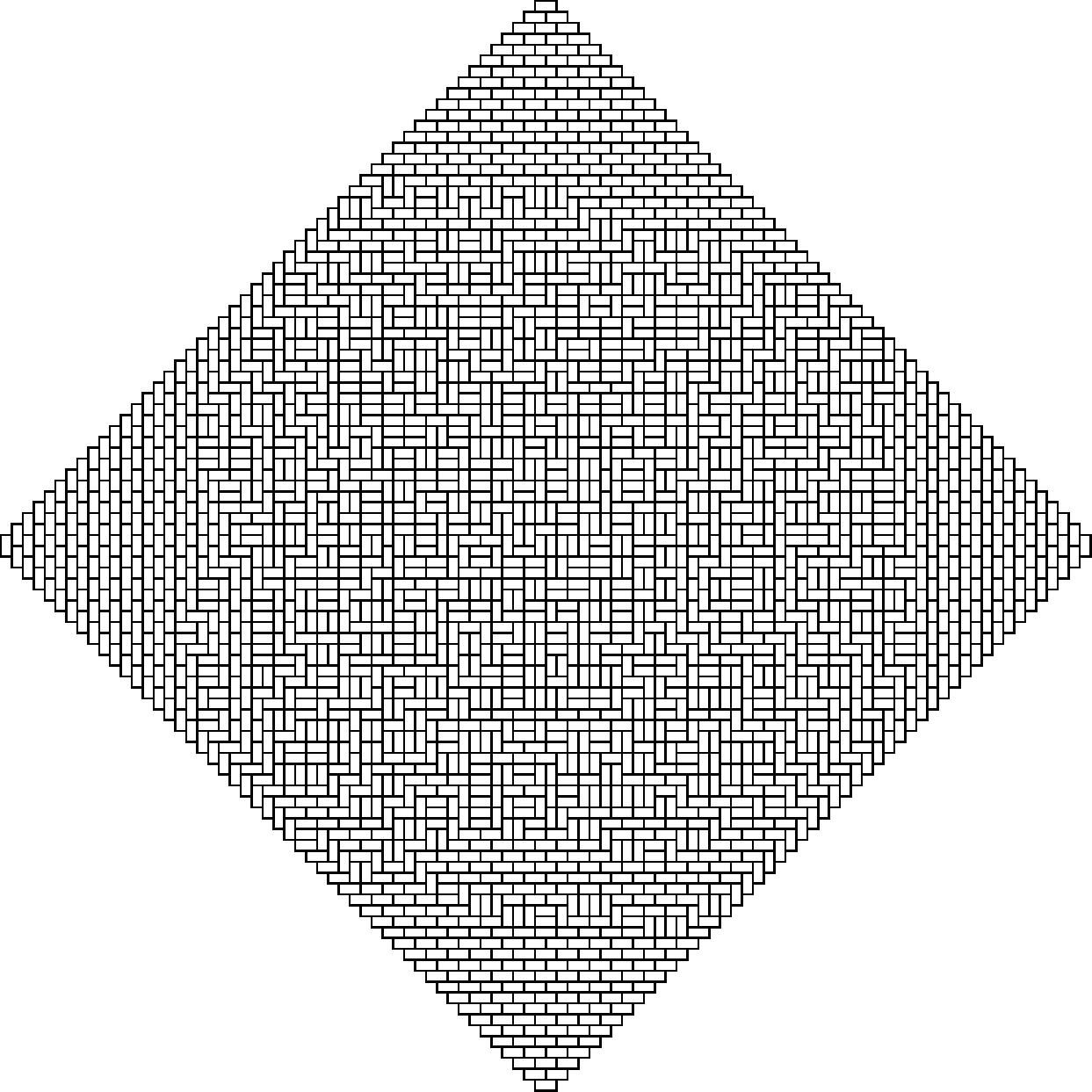}}; 
\node at (2,-5){ \includegraphics[scale = .8]{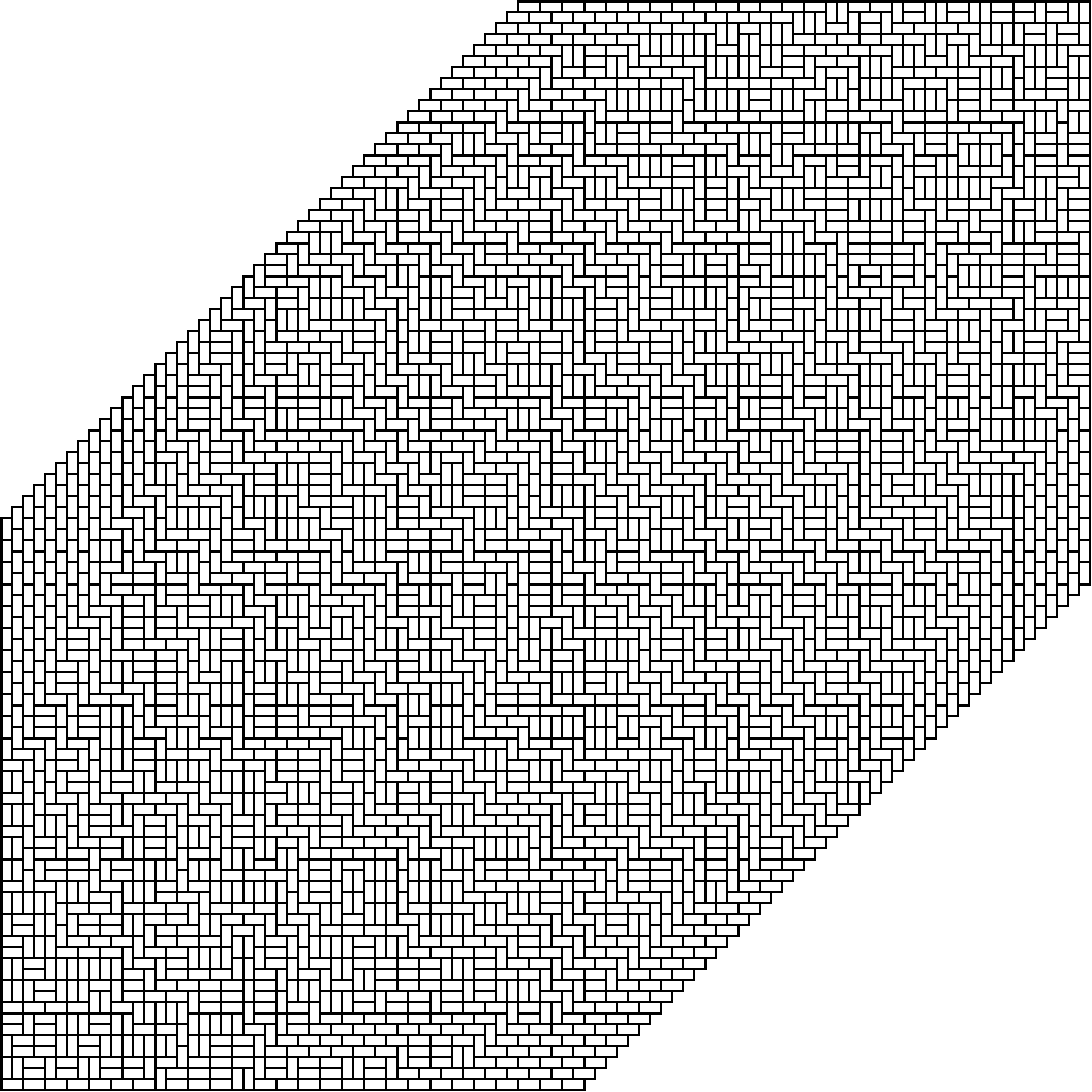}};
\end{tikzpicture}
\caption{Some domino tilings.}
\end{figure}

\begin{PicFigure}[h]
\subfloat[]{
\begin{tikzpicture}
\node at (0,0){\includegraphics[scale = .4]{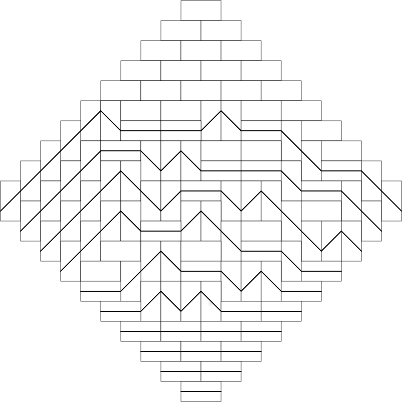}};
\draw[thin, decoration={markings,mark=at position 1 with {\arrow[scale=2]{>}}},
    postaction={decorate}] (0,0) -- (2.6,0); \node at  (2.72,0) { \tiny $ x $ };
\draw[thin, decoration={markings,mark=at position 1 with {\arrow[scale=2]{>}}},
    postaction={decorate}] (0,0) -- (0,2.6); \node at  (0,2.72) { \tiny $ y $ };
\fill (0,.8) circle [radius=.05];
\end{tikzpicture}
} \; \; \; \; \;
\subfloat[]{
\includegraphics[width=0.48\textwidth]{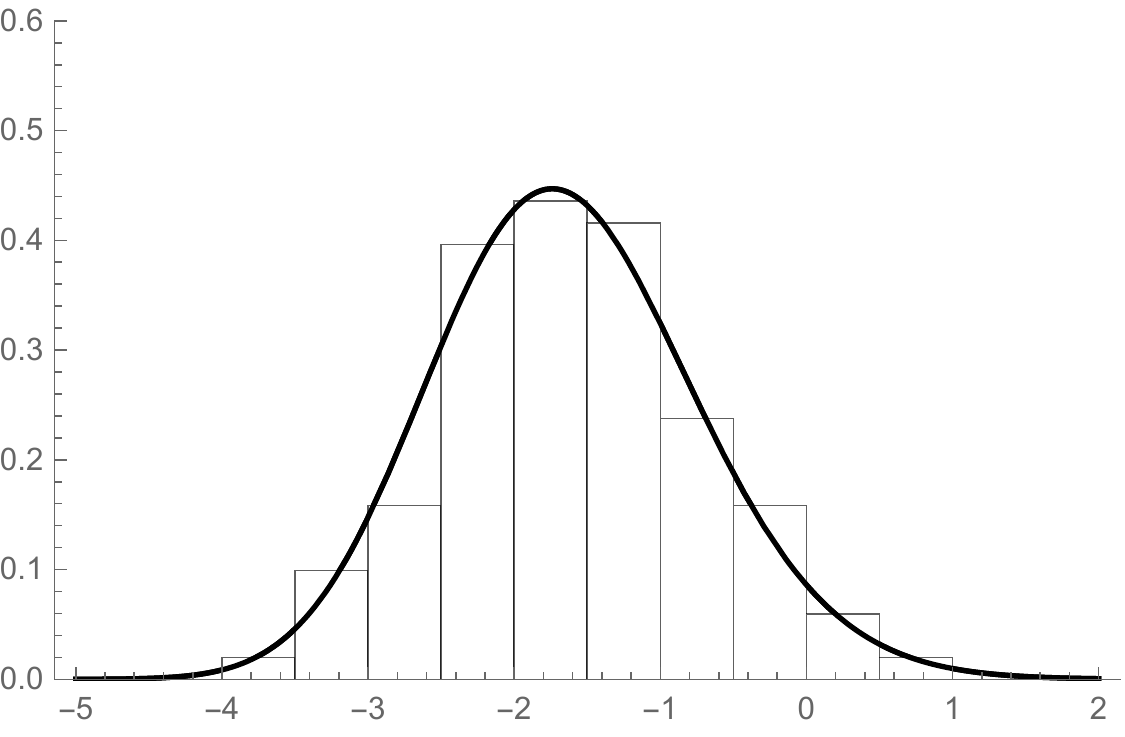}}
\caption{\cite{J} showed that the fluctuations of the top-most path, above which all tiles are horizontal, converges to the Airy process. In particular, the $y$-intercept of the path as shown in (A), after appropriate rescaling, converges to the Tracy-Widom distribution $ F_2 $. (B) shows a normalized histogram of the $ y $-intercept computed from 100 random tilings of an Aztec diamond of size 300, with the distribution $F_2$ superimposed in bold. }
\end{PicFigure}

\begin{figure}[h]
\begin{minipage}[t]{0.45\textwidth}
\begin{center}
\begin{tikzpicture}
\begin{scope}[scale=.5, transform shape]
\node at (0,0){\includegraphics[scale=.5]{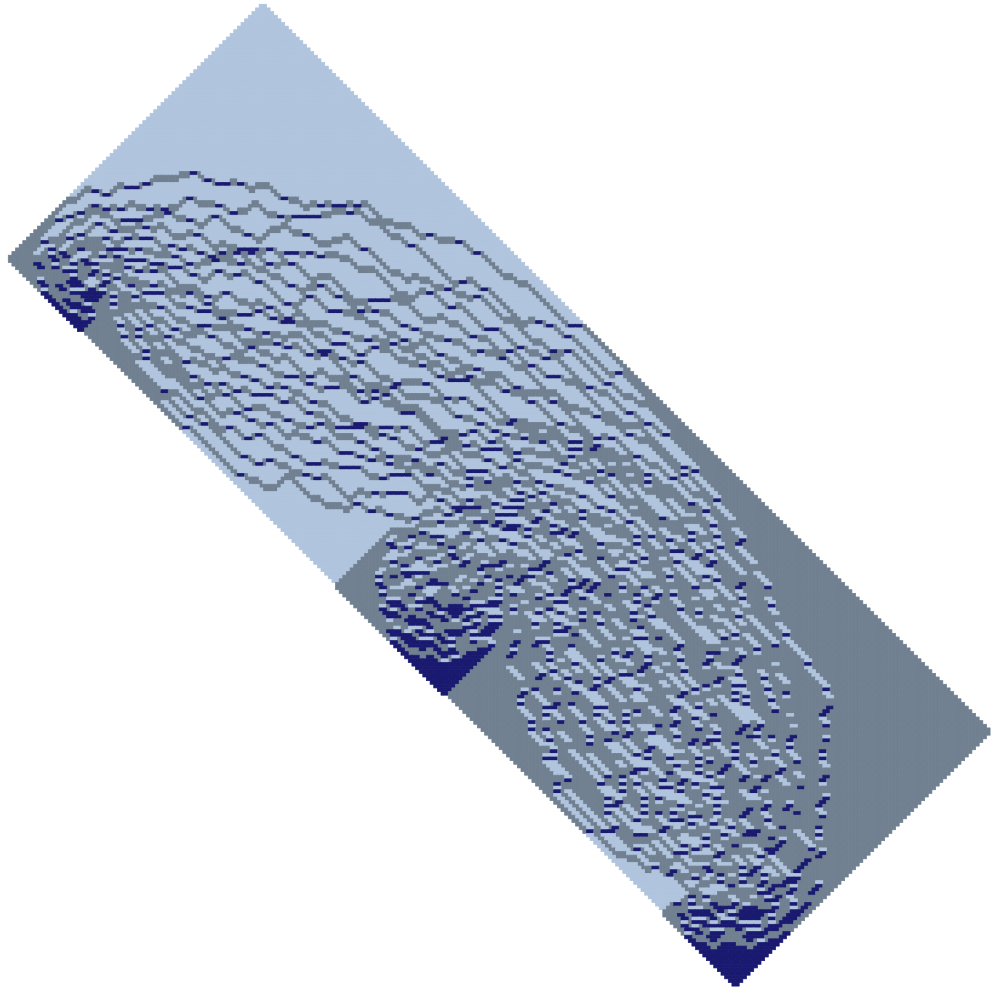}};
\begin{scope}[rotate=-45.0,transform shape]
\node at (0,0){\includegraphics[scale=.89]{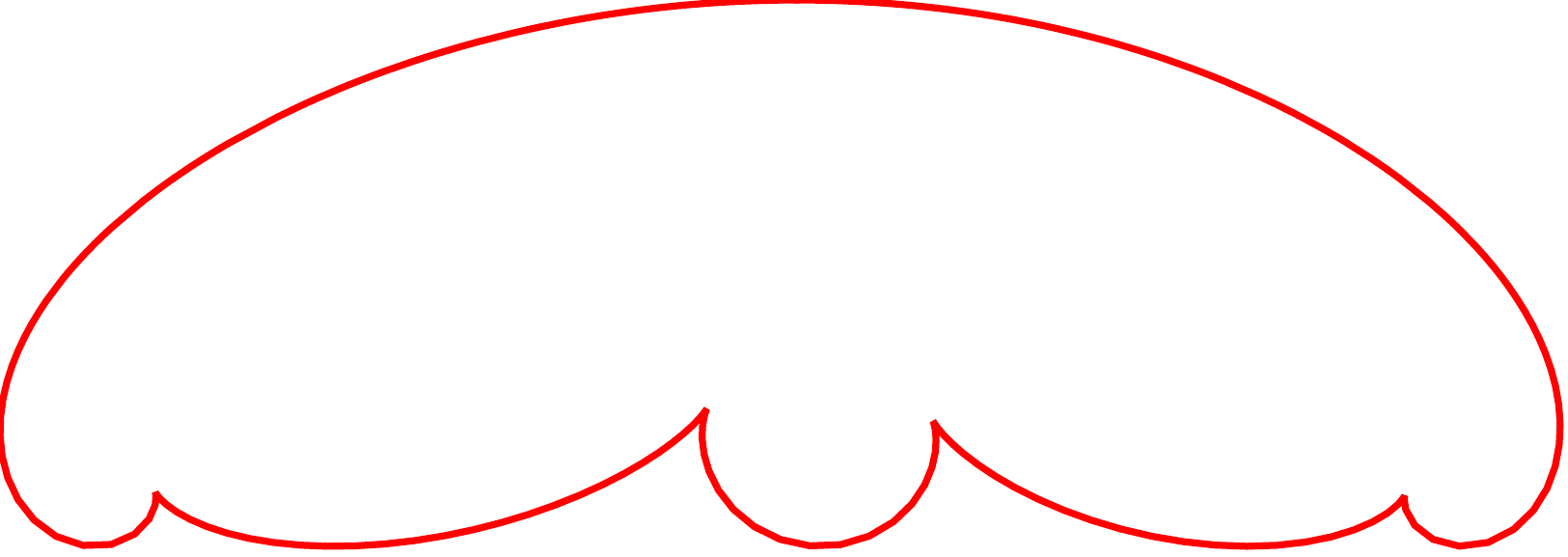}};
\end{scope}
\end{scope}
\end{tikzpicture}   
\captionsetup{width=\linewidth}
    \caption{A tiling of a rectangular Aztec diamond, with the Arctic curve computed by \cite{Knizel} superimposed in red. }
\end{center}
  \end{minipage}
\; \; \;
  \begin{minipage}[t]{0.45\textwidth}
\begin{center}
\begin{tikzpicture}
\node at (0,0){\includegraphics[scale =.25]{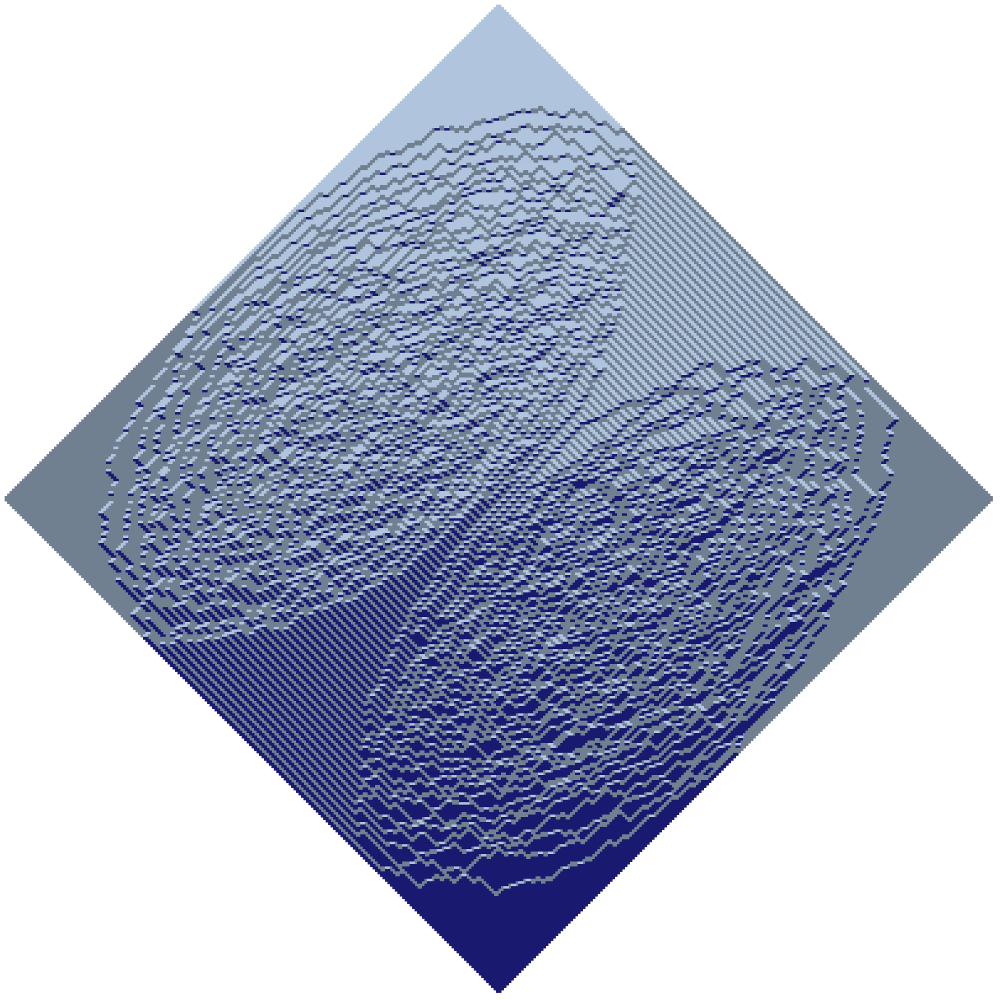}};
\end{tikzpicture}
\captionsetup{width=\linewidth}
\caption{ A random tiling of the Aztec diamond with volume weights (see Section \ref{sec:weights}) $ q = 20 $ for all black vertices and $ q = 1/20 $ for all white vertices. See \cite{Cedric} for details.}
\end{center}
\vfill
  \end{minipage}
\end{figure}

\begin{PicFigure}[h]
\includegraphics[scale = .5]{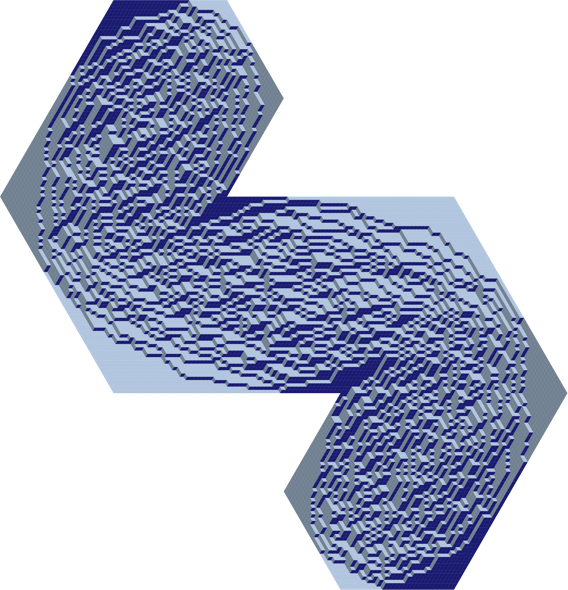}
\caption{A random tiling of a weird region by lozenges.}
\end{PicFigure}

\begin{PicFigure}[h]
\includegraphics[scale = 1.3]{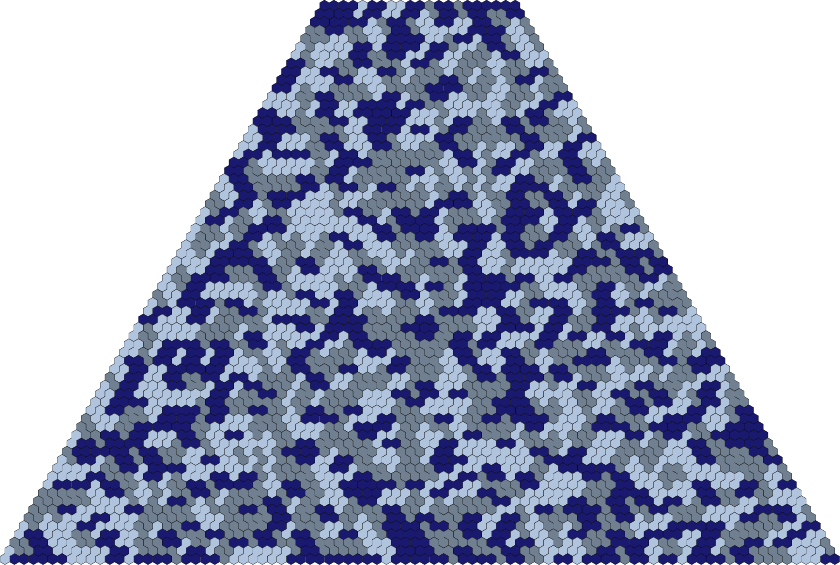}
\caption{A tiling of a trapezoid by bibones. Bibone tilings, which correspond to dimers on the triangular lattice, do not develop Arctic curves or limit shapes.}
\end{PicFigure}

\begin{PicFigure}[h]
\subfloat[Subfigure 1][]{
\includegraphics[width=0.44\textwidth]{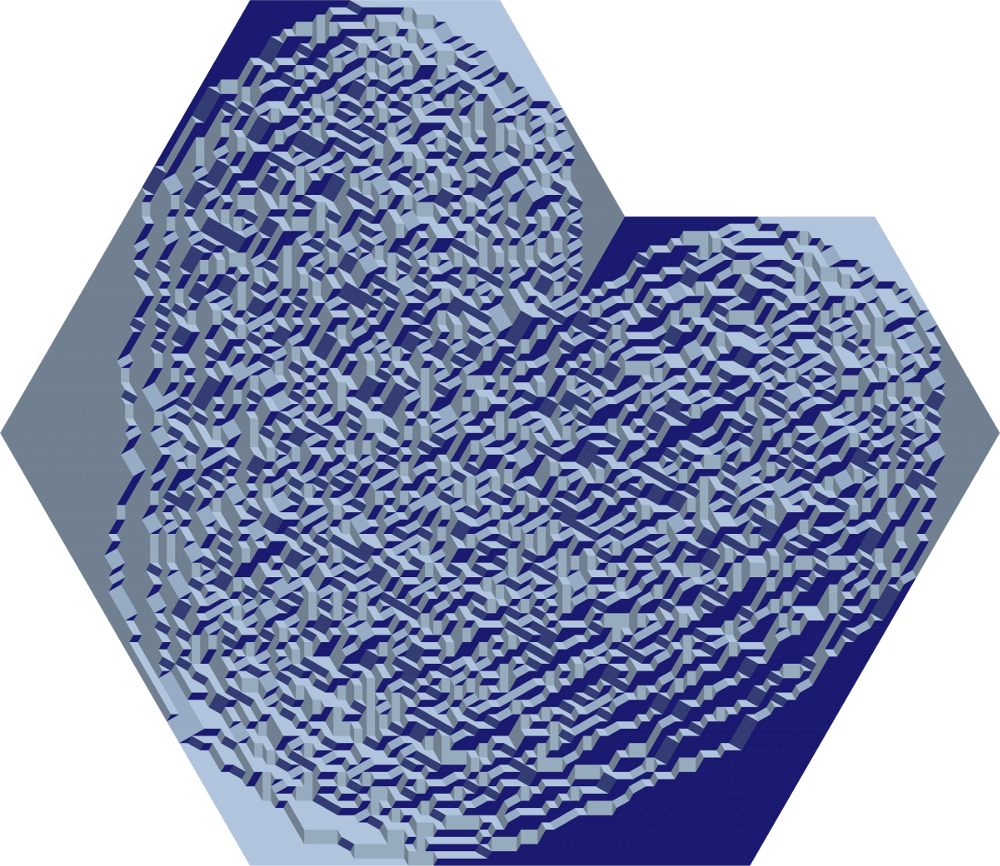}
\label{fig:pa}} \; \; \; 
\subfloat[Subfigure 2][]{
\includegraphics[width=0.44\textwidth]{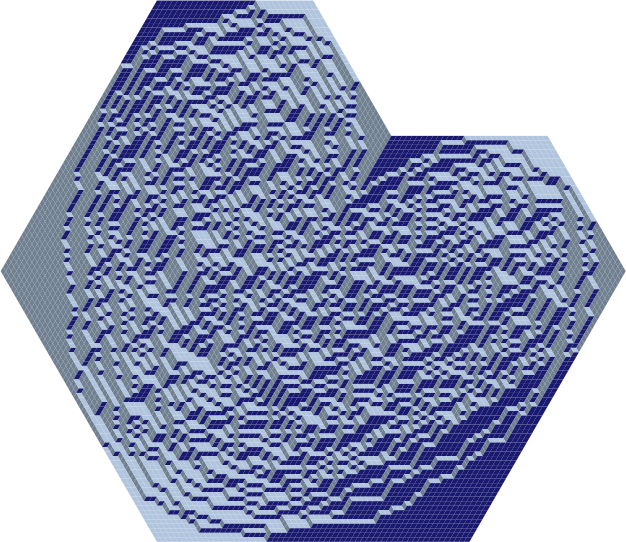}
\label{fig:pb}}
\caption{A tiling of a partial hexagon (A) by rectangles and triangles, and (B) by lozenges.}
\end{PicFigure}
\begin{PicFigure}[h]
\includegraphics[width=0.7\textwidth]{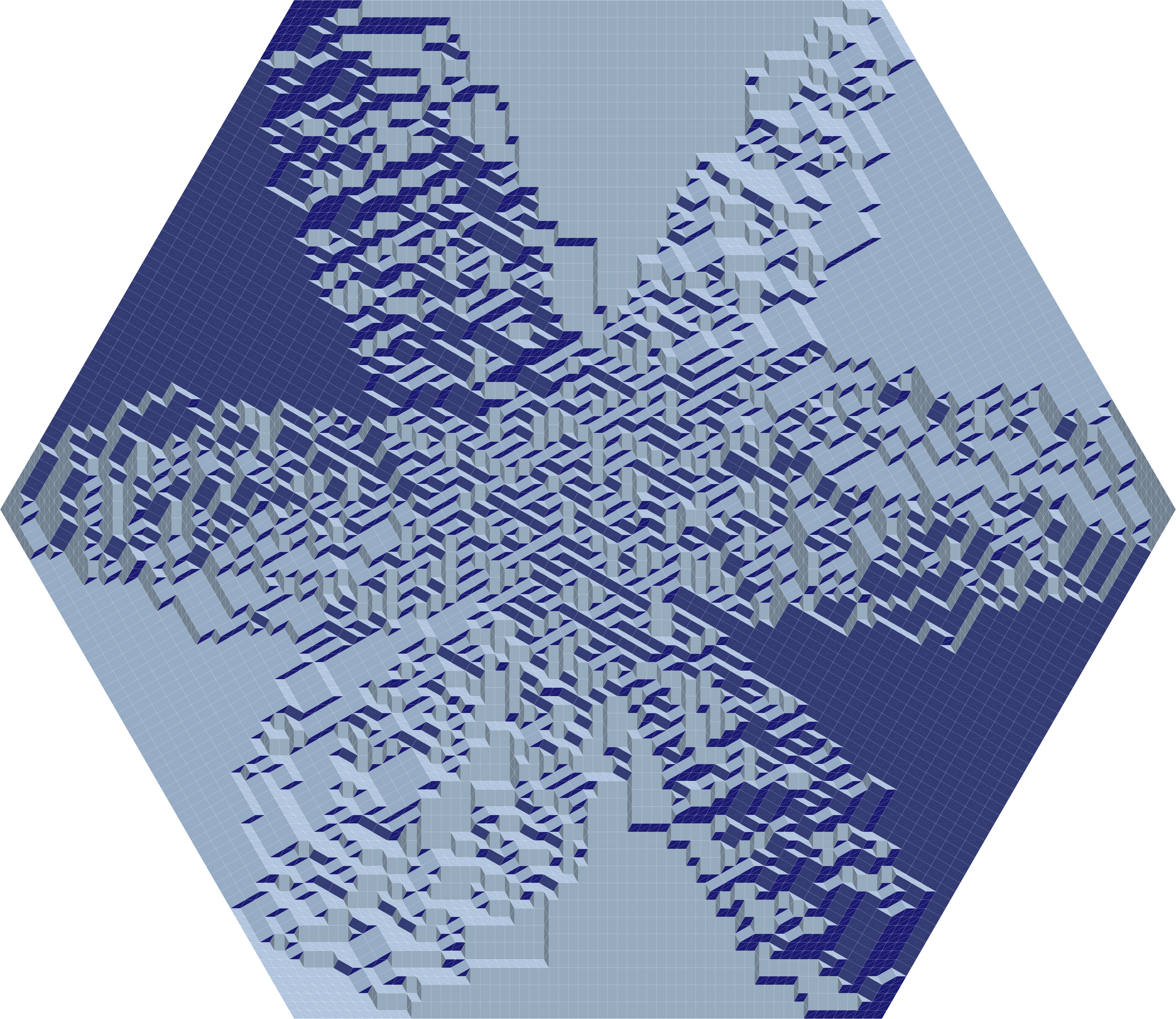}
\caption{ Choosing weights $ t = .5, r = 1, c = 1 $ (see Section \ref{sec:recttriangle}) produces tilings that look like snowflakes. We observed large fluctuations in the boundaries of the arms as compared to arctic curves of lozenge tilings.}
\label{fig:Phase}
\end{PicFigure}

\begin{figure}[h]
\subfloat[]{
\includegraphics[width=0.45\textwidth]{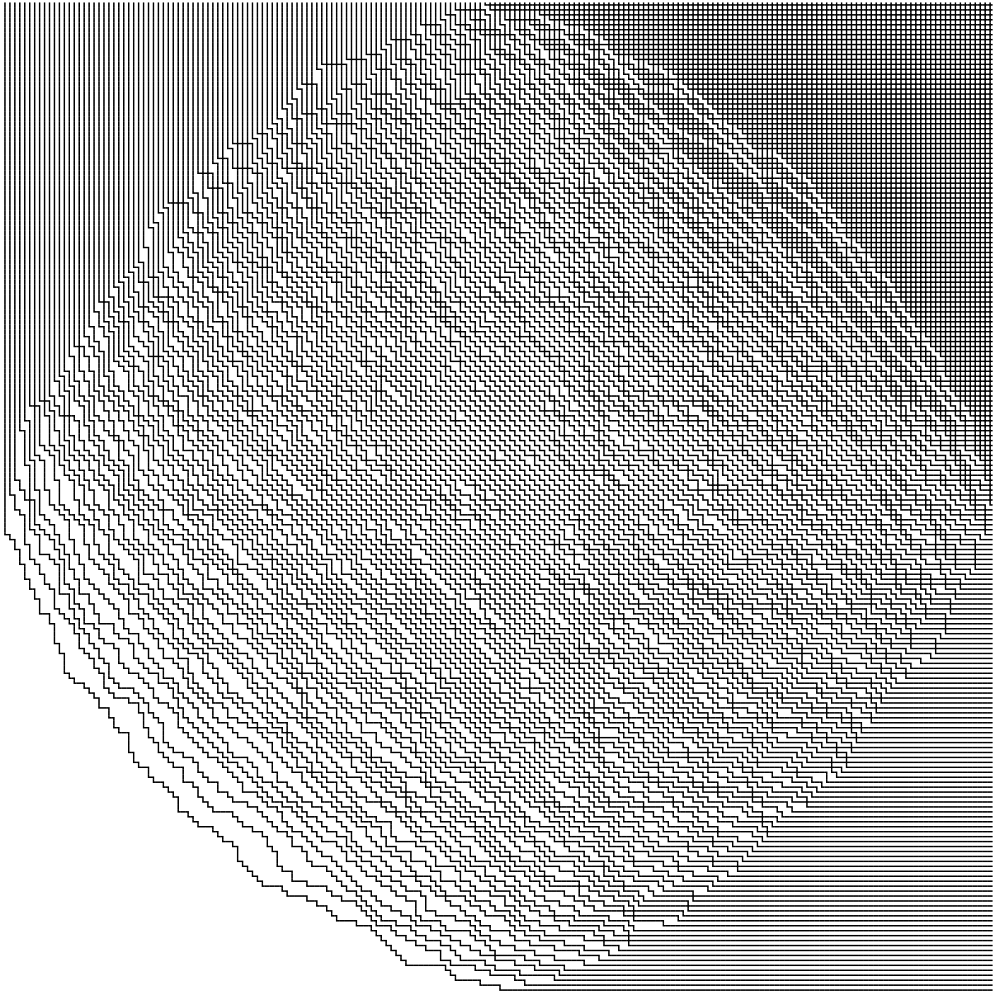}} \; \; \;
\subfloat[]{
\includegraphics[width=0.45\textwidth]{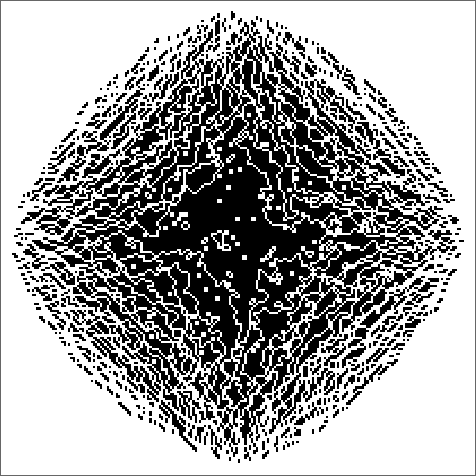}} \\ 
\subfloat[]{
\begin{tikzpicture}
\begin{scope}[yscale=1,xscale=-1, transform shape]
\node at (0,0){ \includegraphics[width=0.45\textwidth]{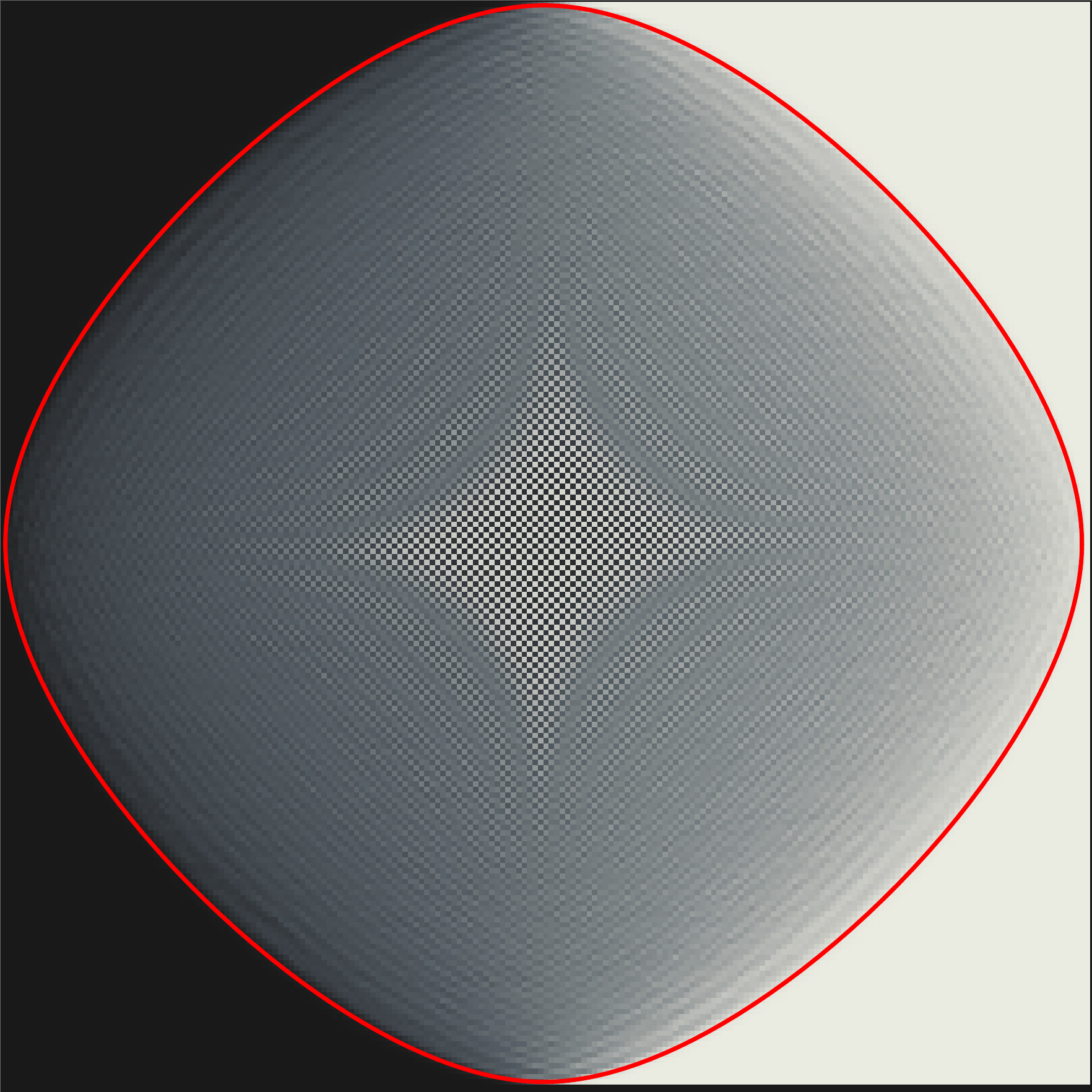}};
\end{scope}
\end{tikzpicture}
} \; \; \;
\subfloat[]{
\includegraphics[width=0.45\textwidth]{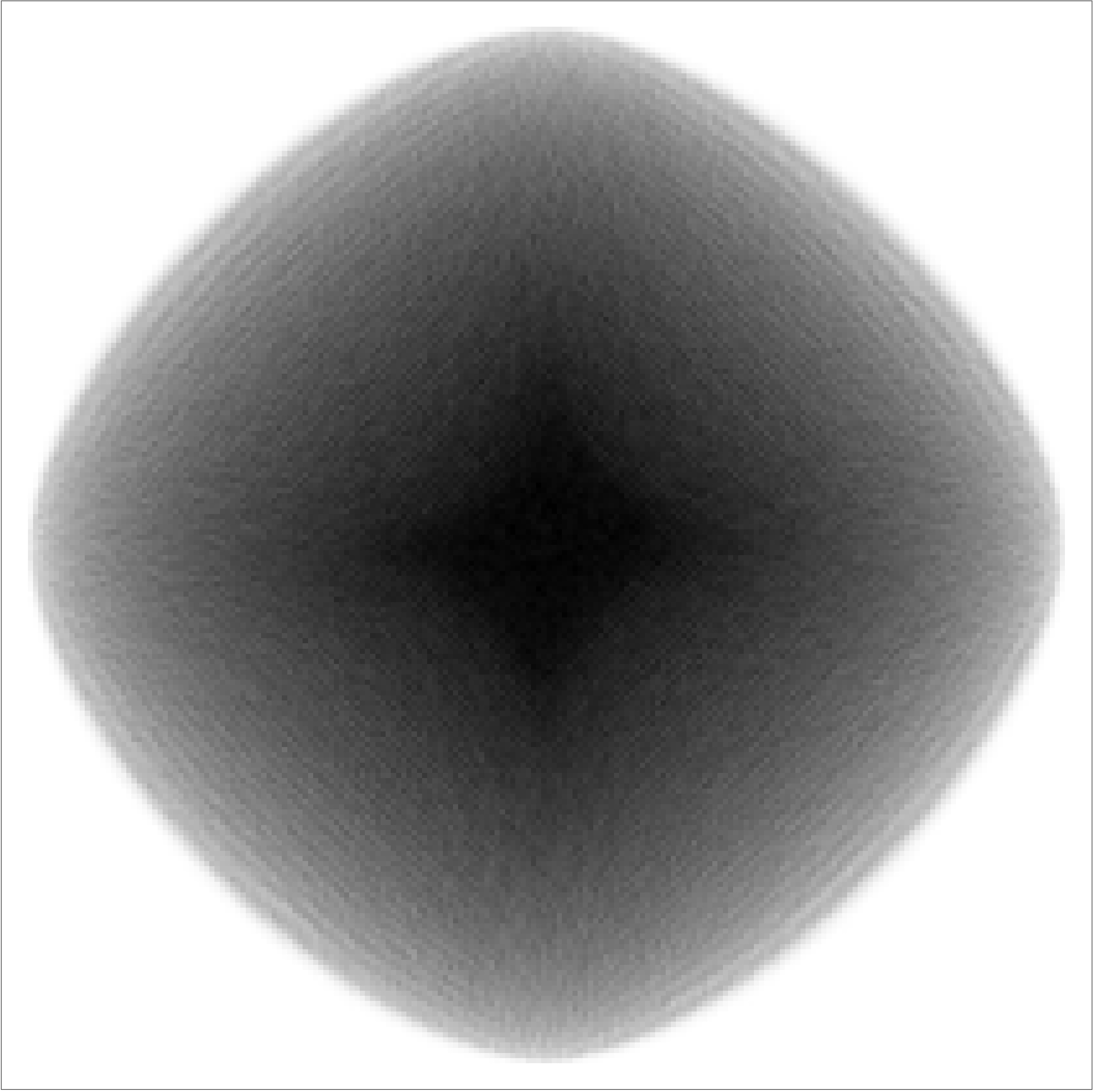}}\\
\subfloat[]{
\begin{tikzpicture}
\begin{scope}[yscale=-1,xscale=-1, transform shape]
\node at (2,-1.2) { \includegraphics[scale = .06]{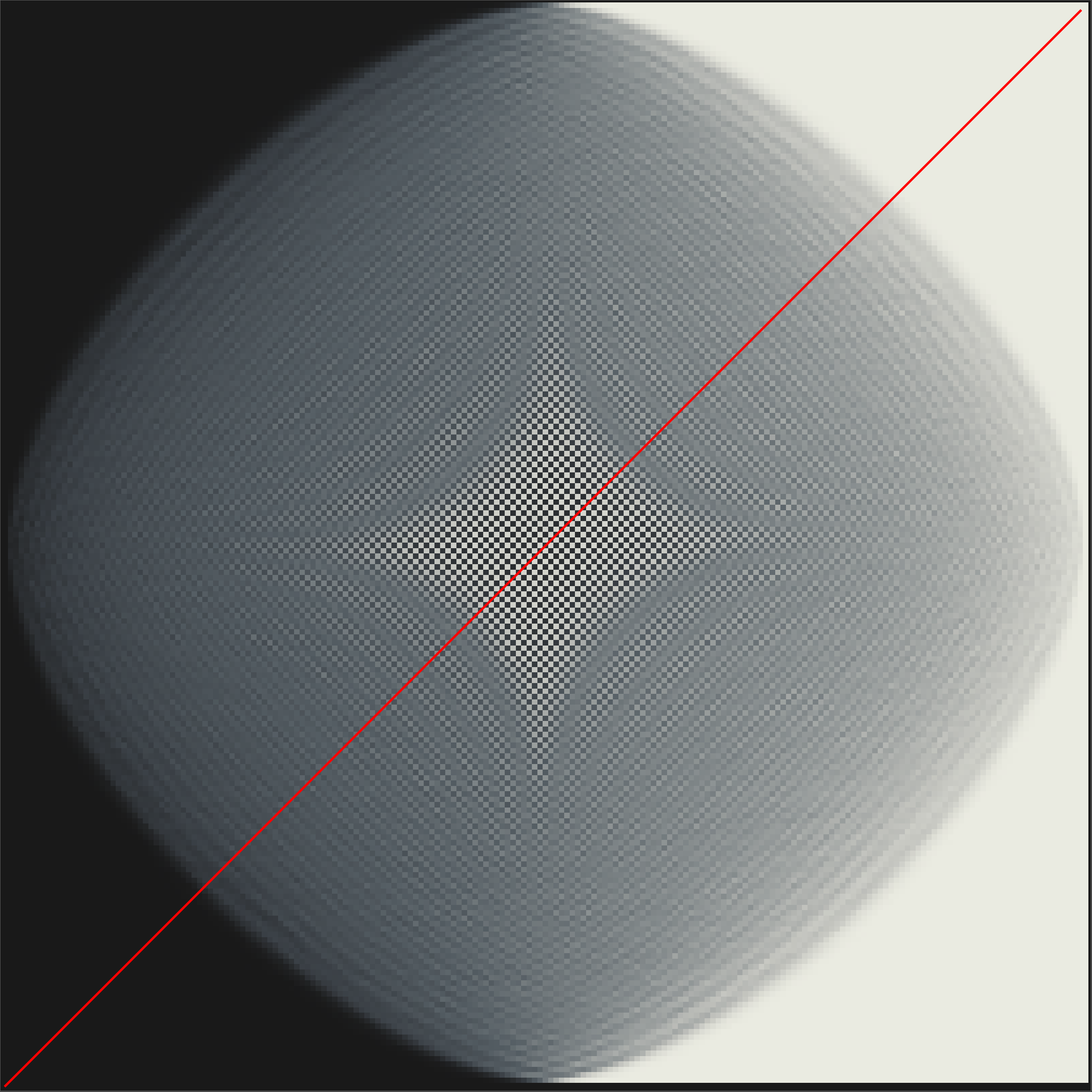}};
\end{scope}
\node at (0,0){ \includegraphics[scale = .4]{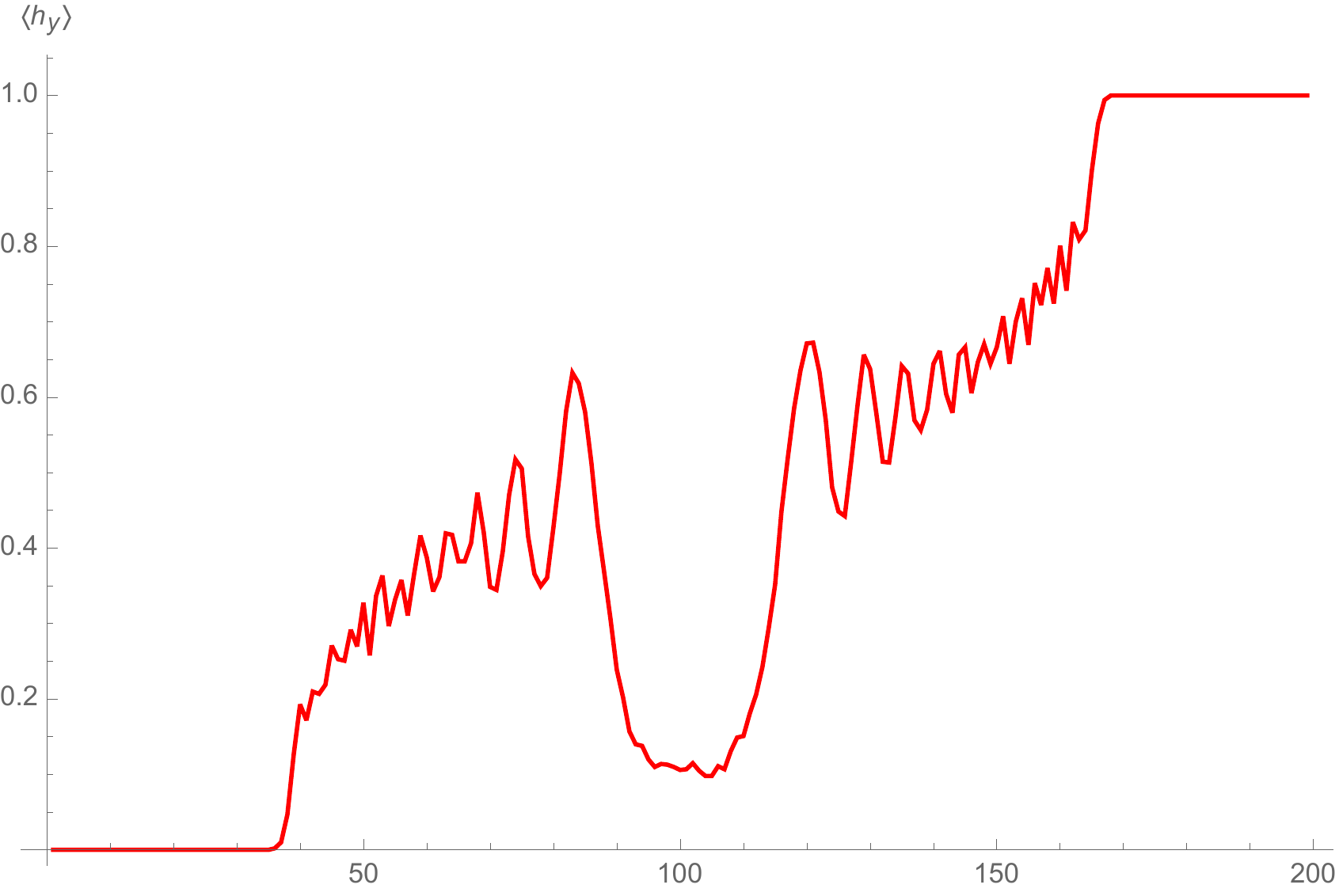}};
\end{tikzpicture}
} \; \; \;
\caption{The six-vertex model with weights $ a = 1, b = 1, c = \sqrt{8} $, ($ \Delta = -3$), in a square with fixed domain wall boundary conditions. (A) shows a random configuration and (B) shows the $ c$-vertices of the random configuration in black. (C) shows the average density of horizontal edges computed, with 1000 random configurations. The arctic curve computed by \cite{ColomoPronko} is superimposed in red. (D) shows the average density of $c$-vertices. (E) shows the cross-section profile of the horizontal edge density along a diagonal slice, which was studied in \cite{korepin}.}
\end{figure}

\begin{figure}[h]
\begin{minipage}[t]{0.45\textwidth}
\begin{center}
\includegraphics[width=\textwidth]{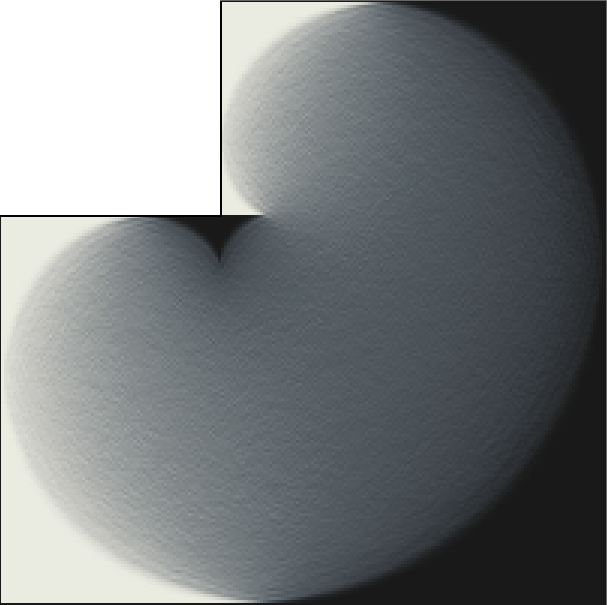}
\captionsetup{width=\linewidth}
    \caption{The average density of horizontal edges in with weight $ \Delta = 0 $ in an L-shaped region with domain wall type boundary conditions, computed with 1000 samples.}
\end{center}
  \end{minipage}
\; \; \;
  \begin{minipage}[t]{0.45\textwidth}
\begin{center}
\includegraphics[width=\textwidth]{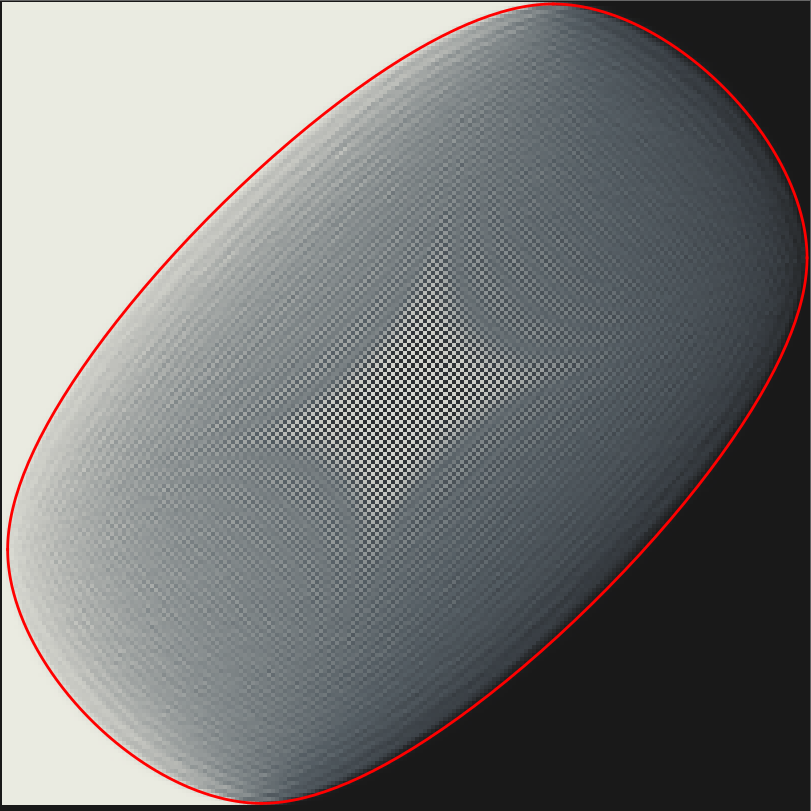}
\captionsetup{width=\linewidth}
    \caption{The average density of horizontal edges with weights $ a = 2b, \; \Delta = - 3 $, computed with 1000 samples. The red curve is the arctic curve computed by \cite{ColomoPronko}.}
\end{center}
\vfill
  \end{minipage}
\end{figure}

\begin{figure}[h]
\subfloat[Subfigure 1][]{
\includegraphics[width=0.48\textwidth]{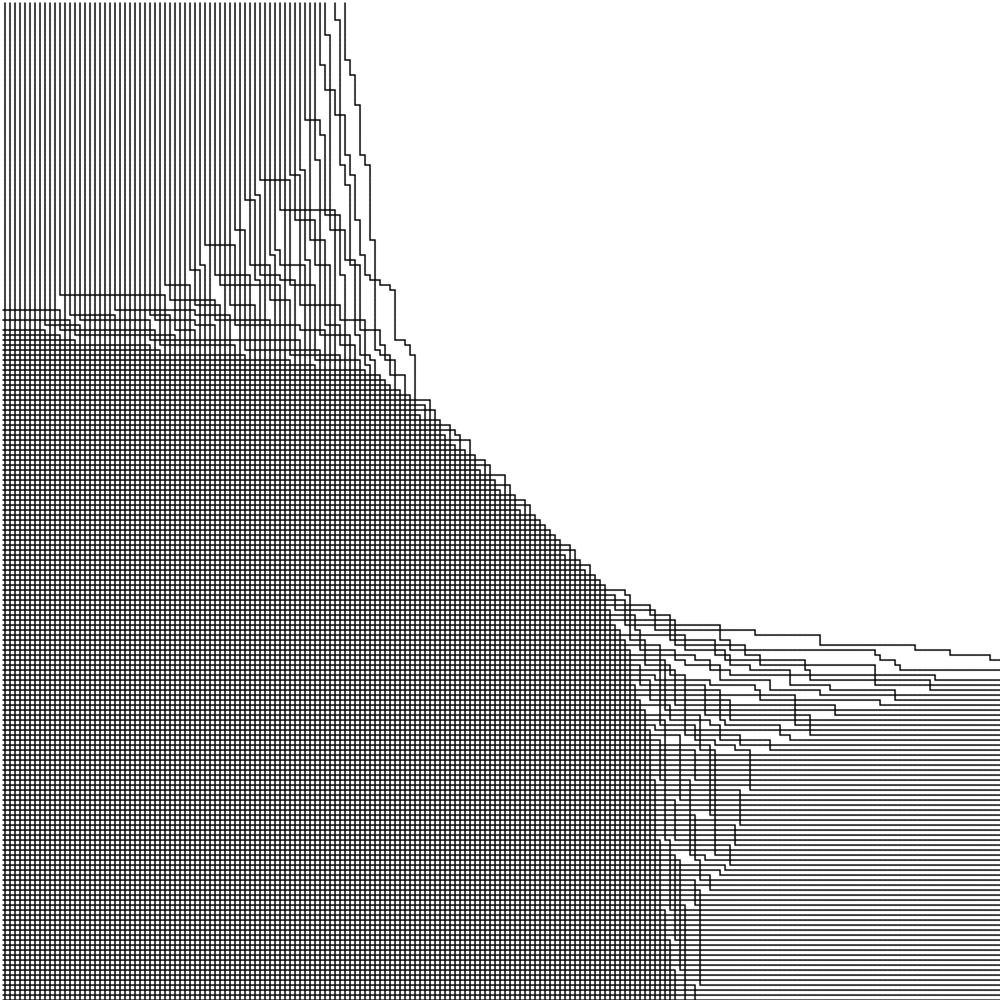}}
\subfloat[Subfigure 2][]{
\includegraphics[width=0.48\textwidth]{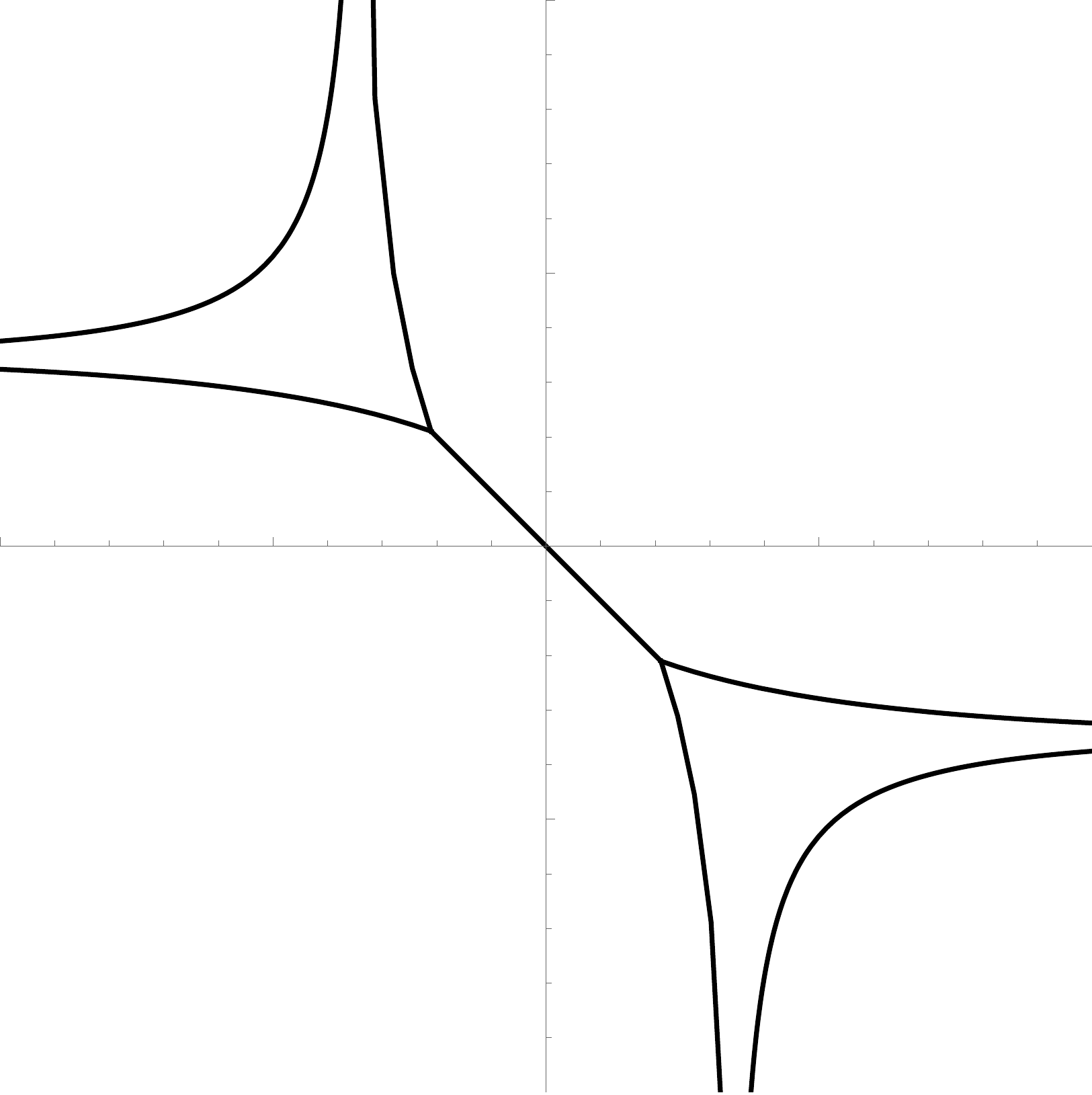}}
\caption{By the Wulff construction, the toroidal free energy $ f(H,V) $ is the limit shape of the volume-constrained model with special boundary conditions. Figure A shows a random configuration with weights $ a = 2, \; b = 1,\; c=.8 $ and volume weights. Figure B shows the free energy phase diagram for the same weights \cite{PR}.}
\end{figure}

\begin{figure}[h]
\begin{minipage}[b]{0.49\textwidth}
\begin{center}
\includegraphics[width=\textwidth]{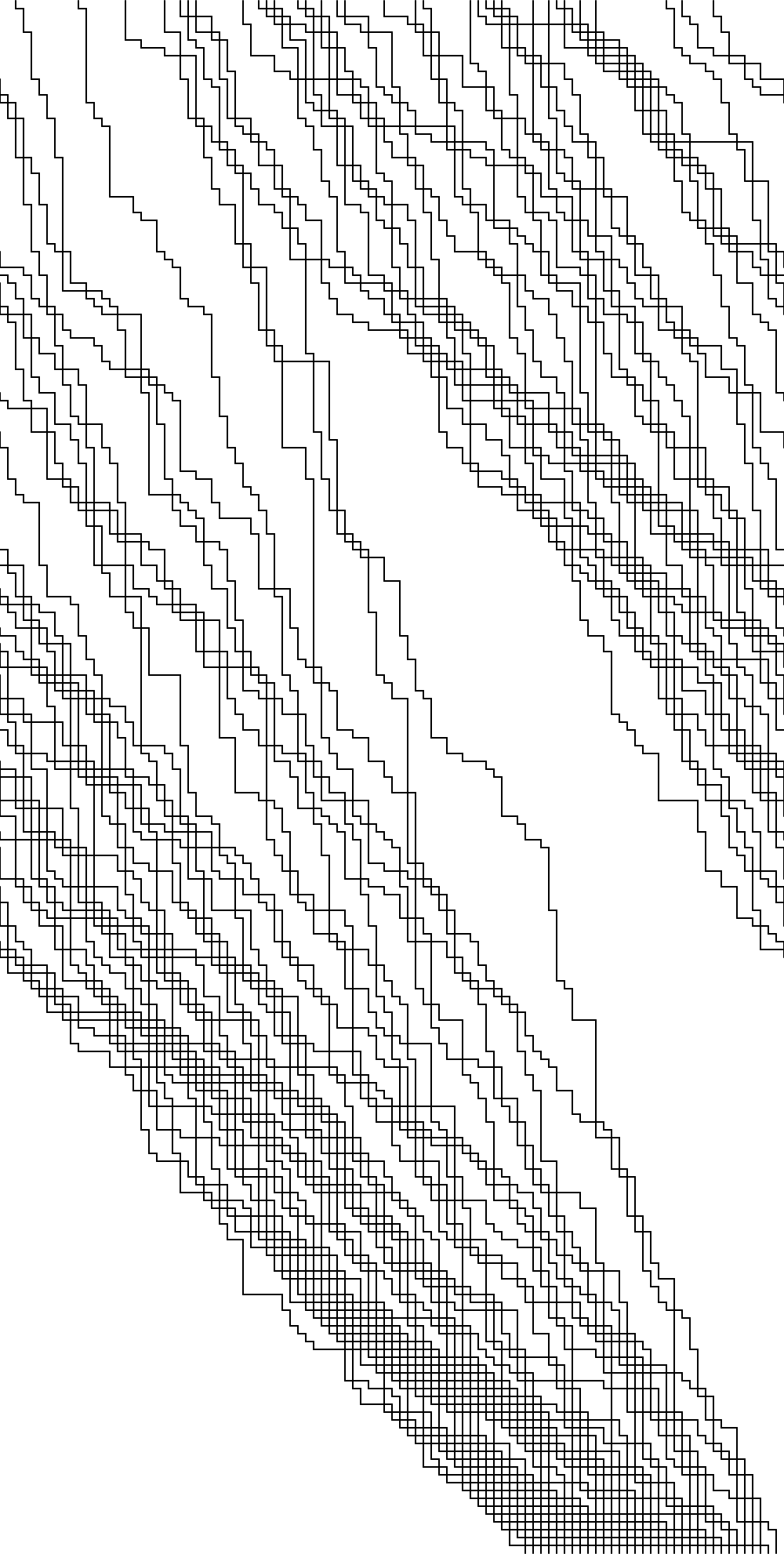}
\end{center}
    \caption*{(A)}

  \end{minipage}
  \begin{minipage}[b]{0.40\textwidth}
\begin{center}
\includegraphics[scale=.6]{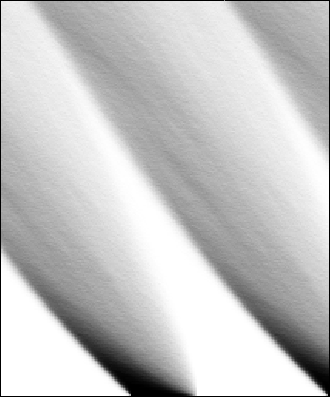}
    \caption*{(B)}
\vspace{20pt}
\input{Sto6VChar}
   \caption*{(C)}
\end{center}
  \end{minipage}

\caption{The six-vertex model at the stochastic point with weights $ ( a_1,a_2,b_1,b_2,c_1,c_2) = (1,1, .3,.7,.3,.7) $ on a cylinder with fixed step boundary conditions at the bottom and free boundary conditions at the top. For details about the stochastic six vertex model, see \cite{GS, BCG, RS}. (A) shows a random configuration on the cylinder. (B) shows the average density of paths, taken over 100 sample, with empty space shaded in white. In the thermodynamic limit, the density of paths is described by a Burgers-type equation that can be solved by characteristics. (C) shows the characteristic lines, with shocks drawn in bold and the rarefaction fans shaded in grey. }
\end{figure}

\end{document}

%% file: Sto6VChar.tex
\begin{tikzpicture}[baseline]
\begin{scope}[shift={(0,.05)}, scale = 3.2]
\begin{scope}
\clip(0,0) rectangle (1.65,2);
\begin{scope}

\draw[white, fill = black!10!white] (1,0) -- (1-0.428571*1.73875,1.73875)--(-.68942+.06,3)--(-1.68942+.06,3)--(-0.428571*1.51875+.06,1.51875)-- (.1-0.428571*1.15052+.06,1.15052)--(.2-0.428571*0.833333+.06,0.833333)--  (.4-0.428571*0.35+.06,0.35) -- (1-2.33333*.26,.26)-- (1,0);

\draw (0,0) -- (-0.428571*1.73875,1.73875);
\draw (.1,0) -- (.1-0.428571*1.34052,1.34052);
\draw (.2,0) -- (.2-0.428571*1.023333,1.023333);
\draw (.3,0) -- (.3-0.428571*0.717188,0.717188);
\draw (.4,0) -- (.4-0.428571*0.46,0.46);
\draw (.5,0) -- (.5-0.428571*0.27,0.27);
\draw (.6,0) -- (.6-0.428571*0.1,0.1);

\draw (.7,0) -- (.7-2.333331*0.025,0.025);
\draw (.8,0) -- (.8-2.33333*0.11,0.11);
\draw (.9,0) -- (.9-2.33333*0.175,0.175);
\draw (1,0) -- (1-2.33333*.26,.26);

\draw [domain = 0:.3, smooth, ultra thick] plot ({.66-\x},{\x});
\draw [domain = .3:1.51875, smooth, ultra thick] plot ({0.771429 +.06 - 0.625969*sqrt(\x) - 0.428571*\x},{\x});
\draw [domain = 1.51875:3, smooth, ultra thick] plot ({.56 - 0.0867857/ \x - 0.720165*\x},{\x});

\begin{scope} \clip (1,0) -- (1-0.428571*1.73875,1.73875)--(-.68942+.06,3)--(-1.68942+.06,3)--(-0.428571*1.51875+.06,1.51875)-- (.1-0.428571*1.15052+.06,1.15052)--(.2-0.428571*0.833333+.06,0.833333)--  (.4-0.428571*0.35+.06,0.35) -- (1-2.33333*.26,.26)-- (1,0);

\draw[gray] (1,0)--(-.6,3);
\draw[gray] (1,0)--(-1,3);
\draw[gray] (1,0)--(-1.4,3);
\draw[gray] (1,0)--(-1.8,3);
\draw[gray] (1,0)--(-2.3,3);
\draw[gray] (1,0)--(-2.8,3);
\draw[gray] (1,0)--(-3.4,3);
\draw[gray] (1,0)--(-4.2,3);
\end{scope}

\end{scope}

\begin{scope}[shift={(1,0)}]
\draw[white, fill = black!10!white] (1,0) -- (1-0.428571*1.73875,1.73875)--(-.68942+.06,3)--(-1.68942+.06,3)--(-0.428571*1.51875+.06,1.51875)-- (.1-0.428571*1.15052+.06,1.15052)--(.2-0.428571*0.833333+.06,0.833333)--  (.4-0.428571*0.35+.06,0.35) -- (1-2.33333*.26,.26)-- (1,0);

\draw (0,0) -- (-0.428571*1.73875,1.73875);
\draw (.1,0) -- (.1-0.428571*1.34052,1.34052);
\draw (.2,0) -- (.2-0.428571*1.023333,1.023333);
\draw (.3,0) -- (.3-0.428571*0.717188,0.717188);
\draw (.4,0) -- (.4-0.428571*0.46,0.46);
\draw (.5,0) -- (.5-0.428571*0.27,0.27);
\draw (.6,0) -- (.6-0.428571*0.1,0.1);

\draw (.7,0) -- (.7-2.333331*0.025,0.025);
\draw (.8,0) -- (.8-2.33333*0.11,0.11);
\draw (.9,0) -- (.9-2.33333*0.175,0.175);
\draw (1,0) -- (1-2.33333*.26,.26);

\draw [domain = 0:.3, smooth, ultra thick] plot ({.66-\x},{\x});
\draw [domain = .3:1.51875, smooth, ultra thick] plot ({0.771429 +.06 - 0.625969*sqrt(\x) - 0.428571*\x},{\x});
\draw [domain = 1.51875:3, smooth, ultra thick] plot ({.56 - 0.0867857/ \x - 0.720165*\x},{\x});

\begin{scope} \clip (1,0) -- (1-0.428571*1.73875,1.73875)--(-.68942+.06,3)--(-1.68942+.06,3)--(-0.428571*1.51875+.06,1.51875)-- (.1-0.428571*1.15052+.06,1.15052)--(.2-0.428571*0.833333+.06,0.833333)--  (.4-0.428571*0.35+.06,0.35) -- (1-2.33333*.26,.26)-- (1,0);

\draw[gray] (1,0)--(-.6,3);
\draw[gray] (1,0)--(-1,3);
\draw[gray] (1,0)--(-1.4,3);
\draw[gray] (1,0)--(-1.8,3);
\draw[gray] (1,0)--(-2.3,3);
\draw[gray] (1,0)--(-2.8,3);
\draw[gray] (1,0)--(-3.4,3);
\draw[gray] (1,0)--(-4.2,3);
\end{scope}
\end{scope}

\begin{scope}[shift={(2,0)}]
\draw[white, fill = black!10!white] (1,0) -- (1-0.428571*1.73875,1.73875)--(-.68942+.06,3)--(-1.68942+.06,3)--(-0.428571*1.51875+.06,1.51875)-- (.1-0.428571*1.15052+.06,1.15052)--(.2-0.428571*0.833333+.06,0.833333)--  (.4-0.428571*0.35+.06,0.35) -- (1-2.33333*.26,.26)-- (1,0);

\draw (0,0) -- (-0.428571*1.73875,1.73875);
\draw (.1,0) -- (.1-0.428571*1.34052,1.34052);
\draw (.2,0) -- (.2-0.428571*1.023333,1.023333);
\draw (.3,0) -- (.3-0.428571*0.717188,0.717188);
\draw (.4,0) -- (.4-0.428571*0.46,0.46);
\draw (.5,0) -- (.5-0.428571*0.27,0.27);
\draw (.6,0) -- (.6-0.428571*0.1,0.1);

\draw (.7,0) -- (.7-2.333331*0.025,0.025);
\draw (.8,0) -- (.8-2.33333*0.11,0.11);
\draw (.9,0) -- (.9-2.33333*0.175,0.175);
\draw (1,0) -- (1-2.33333*.26,.26);

\draw [domain = 0:.3, smooth, ultra thick] plot ({.66-\x},{\x});
\draw [domain = .3:1.51875, smooth, ultra thick] plot ({0.771429 +.06 - 0.625969*sqrt(\x) - 0.428571*\x},{\x});
\draw [domain = 1.51875:3, smooth, ultra thick] plot ({.56 - 0.0867857/ \x - 0.720165*\x},{\x});

\begin{scope} \clip (1,0) -- (1-0.428571*1.73875,1.73875)--(-.68942+.06,3)--(-1.68942+.06,3)--(-0.428571*1.51875+.06,1.51875)-- (.1-0.428571*1.15052+.06,1.15052)--(.2-0.428571*0.833333+.06,0.833333)--  (.4-0.428571*0.35+.06,0.35) -- (1-2.33333*.26,.26)-- (1,0);

\draw[gray] (1,0)--(-.6,3);
\draw[gray] (1,0)--(-1,3);
\draw[gray] (1,0)--(-1.4,3);
\draw[gray] (1,0)--(-1.8,3);
\draw[gray] (1,0)--(-2.3,3);
\draw[gray] (1,0)--(-2.8,3);
\draw[gray] (1,0)--(-3.4,3);
\draw[gray] (1,0)--(-4.2,3);
\end{scope}
\end{scope}
\end{scope}

\draw[ultra thin] (1.65,0) -- (1.65,2);
\draw[ultra thin] (0,0) -- (0,2);
\draw[ultra thin] (0,0) -- (1.65,0);
\draw[ultra thin] (0,2) -- (1.65,2);
\draw [ultra thin, black!30!white, dashed] (1,0)--(1,2);

\end{scope}

\end{tikzpicture}